\documentclass[10pt,journal,final,twocolumn]{IEEEtran}
\IEEEoverridecommandlockouts

\usepackage{amsmath,amssymb,amsfonts}
\usepackage{cite}
\usepackage{graphicx}
\usepackage{epstopdf}

\usepackage{booktabs}
\usepackage{graphicx}
\usepackage{textcomp}
\usepackage{xcolor}
\usepackage{times}
\usepackage{subfigure}
\usepackage{amssymb,amsmath}
\usepackage{acronym}  
\usepackage{balance}

\usepackage{bm}         
\usepackage{algorithm} 
\usepackage{algorithmic}

\usepackage{lipsum}
\usepackage{stfloats}

\usepackage{url}
\usepackage{mathrsfs}
\usepackage{bbm}

\usepackage{amsthm}
\usepackage{enumerate}

\usepackage{hyperref}
\hypersetup{hidelinks, 
colorlinks=true,
allcolors=black,
pdfstartview=Fit,
breaklinks=true}

\theoremstyle{definition}

\theoremstyle{remark}

\acrodef{ofdm}[OFDM]{orthogonal frequency division multiplexing}%
\acrodef{miso-ofdm}[MISO-OFDM]{multi-input single-output orthogonal frequency division multiplexing}%

\acrodef{ris}[RIS]{reconfigurable intelligent surface}%
\acrodef{qos}[QoS]{quality of service}%
\acrodef{idft}[IDFT]{inverse discrete Fourier transform}%
\acrodef{dft}[DFT]{discrete Fourier transform}%
\acrodef{cp}[CP]{cyclic prefix}%
\acrodef{csi}[CSI]{channel state information}%
\acrodef{awgn}[AWGN]{additive white Gaussian noise}%

\acrodef{qcqp}[QCQP]{quadratically constrained quadratic program}%
\acrodef{qp}[QP]{quadratic program}%
\acrodef{bs}[BS]{base station}%
\acrodef{ap}[BS]{base station}%
\acrodef{aps}[APs]{access points}%
\acrodef{qos}[QoS]{quality of service}%
\acrodef{ue}[UE]{user equipment}%
\acrodef{snr}[SNR]{signal-to-noise ratio}%
\acrodef{mmwave}[mmWave]{millimeter-wave}%
\acrodef{snr}[SNR]{signal-to-noise ratio}%

\acrodef{sinr}[SINR]{signal-to-interference-plus-noise ratio}%

\acrodef{ser}[SER]{symbol error rate}%
\acrodef{rc}[RC]{reflection coefficient}%
\acrodef{uavs}[UAVs]{unmanned aerial vehicles}%
\acrodef{mimo}[MIMO]{multiple-input multiple-output}%
\acrodef{noma}[NOMA]{non-orthogonal multiple access}%

\acrodef{ace}[ACE]{adaptive cross-entropy}%
\acrodef{wsr}[WSR]{weighted sum-rate}%
\acrodef{udn}[UDN]{ultra-dense network}%
\acrodef{Udn}[UDN]{Ultra-dense network}%

\def\BibTeX{{\rm B\kern-.05em{\sc i\kern-.025em b}\kern-.08em
    T\kern-.1667em\lower.7ex\hbox{E}\kern-.125emX}}

\begin{document}
\title{Spatio-Temporal Electromagnetic Kernel Learning for Channel Prediction}

\author{\IEEEauthorblockN{Jinke Li, Jieao Zhu, and Linglong Dai,~\emph{Fellow, IEEE}}
\thanks{This work was supported in part by the National Key Research and Development Program of China (Grant No. 2023YFB3811503), in part by the National Natural Science Foundation of China (Grant No. 62325106), and in part by the National Natural Science Foundation of China (Grant No. 62031019).}
\thanks{The authors are with the Department of Electronic Engineering, Tsinghua University, Beijing 100084, China, and also with the Beijing National Research Center for Information Science and Technology (BNRist), Beijing 100084, China. (e-mails: lijk23@mails.tsinghua.edu.cn,  zja21@mails.tsinghua.edu.cn, daill@tsinghua.edu.cn).}
}
\maketitle
\begin{abstract}
 Accurate channel prediction is essential for addressing channel aging caused by user mobility. However, the actual channel variations over time are highly complex in high-mobility scenarios, which makes it difficult for existing predictors to obtain future channels accurately. The low accuracy of channel predictors leads to difficulties in supporting reliable communication. To overcome this challenge, we propose a channel predictor based on spatio-temporal electromagnetic (EM) kernel learning (STEM-KL). Specifically,
inspired by recent advancements in EM information theory (EIT), the STEM kernel function is derived. The velocity and the concentration kernel parameters are designed to reflect the time-varying propagation of the wireless signal. We obtain the parameters through kernel learning. Then, the future channels are predicted by computing their Bayesian posterior, with the STEM kernel acting as the prior. 
To further improve the stability and model expressibility, we propose a grid-based EM mixed kernel learning (GEM-KL) scheme. We design the mixed kernel to be a convex combination of multiple sub-kernels, where each of the sub-kernel corresponds to a grid point in the set of pre-selected parameters. This approach transforms non-convex STEM kernel learning problem into a convex grid-based
problem that can be easily solved by weight optimization. Finally, simulation results verify that the proposed STEM-KL and GEM-KL schemes can achieve more accurate channel prediction. This indicates that EIT can improve the performance of wireless system efficiently. 

\end{abstract}


\section{Introduction} 
In recent years, with the development of new applications such as digital twins and virtual reality, the demand for spectral efficiency is predicted to increase rapidly~\cite{wang2024tutorial}. As a key technology in current wireless communication, massive multiple-input multiple-output (MIMO) can achieve significant improvements in spectral efficiency and system capacity~\cite{rusek2012scaling,ngo2013energy,lu2014overview}.

The effective communication of massive MIMO system highly relies on accurate and timely channel state information (CSI). However, dynamic environments, characterized by user mobility, complicate the acquisition of CSI. According to the current 5G standard~\cite{3gpp_38331}, in time-division duplexing (TDD) mode, CSI acquisition, or channel estimation, is performed periodically. When user mobility is high, significant channel changes may occur within a single channel estimation period, leading to outdated CSI. This phenomenon is termed as {\it channel aging}~\cite{zheng2021impact}. For example, when the user speed is $60\,{\rm{km/h}}$, channel aging could result in approximately $30\%$ loss in achievable sum-rate performance~\cite{jiang2022accurate}. 

In future 6G scenarios, as the number of antennas in MIMO systems increases significantly, the number of pilots required for channel estimation will also increase. Although pilot density can be increased to accommodate this growing demand, when the number of antennas increases several times, the pilot density cannot withstand the dramatic increase subsequently. Consequently, extending the channel estimation period becomes inevitable, leading to more severe channel aging. Therefore, addressing channel aging has become an urgent priority for the wireless communication industry.

\subsection{Prior Works}\label{sec:intro prior}
To address the challenges posed by channel aging, various channel prediction techniques have emerged that utilize the {\it temporal correlation} between historical CSI and future CSI. Existing channel prediction methods can be categorized into two main types: Sparsity-based methods, and autoregressive (AR)-based methods.

    {\bf{Sparsity-based methods}} typically exploit the Doppler domain sparse structure of channel responses to predict future channels. For instance, the sum-of-sinusoids model-based predictor~\cite{pena2013estimation} represents the channel response as a combination of sinusoidal waves. This scheme first identifies the dominant sinusoidal components and then use harmonic retrieval method~\cite{wong2008sinusoidal} to obtain these components for channel prediction. In order to be more suitable for predicting massive MIMO channels with a larger number of vector elements, the authors of~\cite{yin2020addressing} proposed the Prony vector (PVEC) method which fits a linear prediction model to the observed channel response. Specifically, PVEC is applicable to predicting uniformly sampled signals composed of damped sinusoidal components. It models the future channel as a linear combination of the past channels, where the combination weights are computed from the received pilot signals. The authors of~\cite{shi2021compressive} believes time-varying channels have sparsity in the Doppler frequency domain. Consequently, compressive sensing algorithms such as Orthogonal Matching Pursuit (OMP)~\cite{lee2016channel} can be used to obtain the dominant Doppler frequencies for predicting future channels. 

{\bf{AR-based methods}} use autoregressive principle to process channel time series~\cite{baddour2005autoregressive}. The original AR prediction method models the future channel as a weighted sum of its past values, where the weights, i.e., the AR parameters, are obtained from the autocorrelation function of channels at different times~\cite{yuan2020machine,chen2012autoregressive,yusuf2021autoregressive}. The Wiener channel predictor and Kalman channel predictor are extensions of AR prediction method~\cite{loschenbrand2023spectral,papazafeiropoulos2015deterministic,lindbom2001tracking,qin2023eigenvector}. The Wiener predictor enhances channel prediction by predicting an autoregressive multivariate random process using a Wiener linear filter~\cite{truong2013effects}. Moreover, the authors of~\cite{kashyap2017performance} and~\cite{kim2020massive} explores the application of the Kalman predictor within a time-correlated channel aging model. This method implements channel prediction by modeling the channel as a linear dynamic system with state and observation equations. It predicts the next state based on the current estimate and the state transition model, then improves this prediction using new CSI to correct the estimate and reduce uncertainty.



The existing two categories of channel prediction methods mentioned above can fulfill channel prediction for massive MIMO systems. However, simply modeling time-varying channels as sinusoidal or Gaussian random processes is inaccurate. Due to inaccurate channel modeling, these methods cannot accurately predict the channel. The low accuracy of channel predictors can lead to difficulties in supporting reliable communication in high-mobility scenarios. Therefore, it is essential to investigate a more accurate channel prediction method.

\subsection{Our Contributions} \label{sec:intro contr}

In order to design a high-accuracy channel predictor, we propose a channel prediction scheme based on {\it electromagnetic kernel learning}, which simultaneously utilizes the spatio-temporal electromagnetic correlation characteristic of the channel from the perspective of electromagnetic information theory (EIT). The contributions of this paper are summarized as follows: 
\begin{itemize}
	\item Unlike existing channel prediction schemes, we use the EIT-based channel model. Inspired by the spatial correlation function based on electromagnetic (EM) physical principles~\cite{zhu2024benefits}, we consider the time-varying property of the channel and derive the spatio-temporal electromagnetic (STEM) correlation function, i.e., STEM kernel. Specifically, we introduce the velocity parameter in the correlation function to describe the user mobility. This STEM kernel originates from EM physics, thus it is more suitable for modeling practical wireless propagation environment than other kernel functions. 
	\item Since the proposed STEM kernel chacterizes the channel temporal correlation, we utilize the STEM kernel to construct time-domain channel predictors. To get the STEM kernel parameter, we formulate a maximum likelihood (ML) problem, where the kernel parameters are optimized to fit the noisy channel observations. Furthermore, we design the velocity and the concentration kernel parameters to reflect the time-varying propagation of the wireless signal. After determining the kernel parameters, the future channels are predicted by computing their Bayesian posterior, with the STEM kernel acting as the prior. Therefore, we introduce EM information into the channel predictor in a physically interpretable way.
	\item To deal with the non-convexity of the ML problem, we convert it into a convex problem by introducing additional grid weight parameters, leading to a convex grid-based problem that can be easily solved by weight optimization. Specifically, the STEM kernel is approximated by a new grid-based EM mixed (GEM) kernel, which is composed of STEM sub-kernels. For each sub-kernel, parameters are fixed at a set of pre-selected grid points, leaving only the weights to be optimized. Thus, the original continuous parameter optimization problem is converted into a discrete weight optimization problem with favorable convexity and reduced complexity. 
	\item Finally, through performance analysis and numerical experiments, it can be verified that the proposed GEM-KL channel predictor outperforms the PVEC and AR baselines, which demonstrates that EIT can benefit the performance of wireless communication systems.
\end{itemize}

\subsection{Organization and Notation}\label{sec:intro org}
The rest of the paper is organized as follows. Section~\ref{sec: sys} introduces the channel model and signal model. Section~\ref{sec: Problem Formulation} formulates the channel prediction problem. In section~\ref{sec: Proposed}, we first introduce electromagnetic correlation function (EMCF). Then, Gaussian process regression (GPR) is introduced for solving channel prediction problem. Kernel learning is considered to improve EM based GPR channel prediction, and finally the GEM-KL scheme is proposed. Simulation results are provided in Section~\ref{sec: sim}, and we conclude this paper in Section~\ref{sec: conclusion}.

\textit{Notations}: ${\bf{X}}$ and ${\bf{x}}$ respectively denote matrices and vectors. $\mathbb{E}[X]$ denotes the expectation of random variable $X(\omega)$; $\mathbb{C}$ denotes the set of complex numbers and $\mathbb{R}$ denotes the set of real numbers; $(\cdot)^\ast $ denotes the conjugate operation; ${[\cdot]^{-1}}$, ${[\cdot]^{\mathsf{T}}}$, ${[\cdot]^{\mathsf{H}}}$ and ${\rm diag}(\cdot)$ denote the inverse, transpose, conjugate-transpose and diagonal operations, respectively; ${\rm i}$ denotes the imaginary unit; $\mathbf{I}_{N}$ is an $N\times N$ identity matrix; For ${\bf{x}}\in\mathbb{C}^{n}$ or $\mathbb{R}^{n}$, $\left | {\bf{x}} \right | =\sqrt{{\bf{x}}^{\mathsf{T}}{\bf{x}}} \in\mathbb{C}$ denotes the pseudonorm; $\left \| {\bf{x}}  \right \|  $ denotes the standard vector 2-norm $ \sqrt{{\bf{x}}^{\mathsf{H}}{\bf{x}}} \in\mathbb{R}_{\ge 0}$; $\hat{\bf{x}}$ denotes ${\bf{x}}/\left | {\bf{x}} \right |$; $\mathfrak{R} \left \{ \cdot  \right \} $ and $\mathfrak{I} \left \{ \cdot  \right \} $ respectively represent the real and imaginary part of the arguments; $j_{m}(x)$ is the $m$th-order spherical Bessel function of the first kind. 

\section{System Model}\label{sec: sys}
In this section, we first review the Gaussian random field (GRF)-based channel model, and then explain the signal model.
\subsection{Channel model}\label{sec:Channel model}
  Traditional channel models express channel matrices as a weighted Gaussian mixture of steering vectors, which is a discrete special case of a Gaussian random field. 
  To capture the continuous varying property of the wireless channel, in this section, we model the channel with a complex symmetric Gaussian random field (CSGRF). 
  Let function ${{h}}\left ( {\bm{\rho}}  \right ) :\mathbb{R}^{4} \to \mathbb{C}$ represent a circularly symmetric Gaussian random field (CSGRF). 
  The variable is ${\bm{\rho}} =({\bf{x}},t  )$, where ${\bf{x}}=(x,y,z)$ represents the spatio location, $t$ represents time indicator, 
  and $({\bf{x}},t)\in \mathbb{R}^{4}$. For any $Q$ points, the joint distribution of their function values $({{h}}({\bm{\rho}}_{1} ),{{h}}({\bm{\rho}}_{2} ),\dots,{{h}}({\bm{\rho}}_{Q} ))$ follows a multivariate Gaussian distribution, then the random field is a Gaussian random field, 
  denoted as ${{h}}({\bm{\rho}})\sim \mathcal{GRF} (0, {k}({\bm{\rho}},{\bm{\rho}}{'} ) )$, and its probability measure is determined by their autocorrelation function

\begin{equation}
\label{eq: autocorrelation function}
\begin{aligned}
{k} ({\bm{\rho}},{\bm{\rho}{'}}) = \mathbb{E}\left [ {{h}} ({\bm{\rho}}){{h}^{\ast}}({\bm{\rho}}{'}) \right ].
\end{aligned}
\end{equation}

The autocorrelation function is usually called the kernel, note that the kernel function of the GRF must be semi-positive definite.
To enable CSGRF to represent the wireless channel, some restrictions should be imposed on ${k} ({\bm{\rho}},{\bm{\rho}{'}}) $ so that the ${{h}}({\bm{\rho}})$ generated by it satisfies the EM propagation constraints. We use ${{h}}({\bm{\rho}})$ to model the electric field distribution ${\bf{E}}({\bm{\rho}}):\mathbb{R}^{4}\to\mathbb{C}^{3}$. Then, the autocorrelation function can be defined as ${\bf{K}}_{\bf{E}} ({\bm{\rho}},{\bm{\rho}{'}}) = \mathbb{E}\left [ {\bf{E}} ({\bm{\rho}}){\bf{E}} ({\bm{\rho}}{'})^{\mathsf{H}}\right ]\in\mathbb{C}^{3\times3}$~\cite{zhu2024electromagnetic}. Similarly, for a channel vector with $N_{\rm{BS}}$ components, it can also be modeled using CSGRF by constructing the autocorrelation function of ${\bm{\rho}}_{n}$ for $n=1,2,\dots,N_{\rm{BS}}$.
\\

\subsection{Signal Model}
 For signal model, a XL-MIMO system is considered, in which a single base station (BS) with $N_{\rm{BS}}$ antennas serves a single user with 1 antenna. 
We will try to solve the problem of uplink channel prediction in a narrowband system. Consider the simplest communication scenario, assuming we use an $N_{\rm{BS}}$-antenna base station with fully digital precoding, where each antenna is connected to a dedicated radio frequency (RF) chain. The uplink signal model is
\begin{equation}
\label{eq: uplink signal model}
\begin{aligned}
{\bf{y}}_{t} &= {\bf{h}}_{t}+{\bf{n}}_{t},\\
\end{aligned}
\end{equation}
where ${\bf{y}}_{t}\in \mathbb{C}^{N_{\rm{BS}} \times 1} $is the BS received pilots at time $t$, ${\bf{h}}_{t}\in \mathbb{C}^{N_{\rm{BS}} \times 1} $is the normalized channel vector satisfying $\mathbb{E}[  \left \| {\bf{h}}_{t} \right \| ^{2}  ] = N_{\rm{BS}} $, and ${\bf{n}}_{t}$ is the complex-valued additive white Gaussian noise (AWGN) with zero mean and covariance $\frac{1}{{{\mathsf{SNR}}}}  {\bf{I}}_{N_{\rm{BS}}} $. The symbol ${\mathsf{SNR}}$ represents the received signal-to-noise ratio of the BS.

We use the least squares (LS) and minimum mean square error (MMSE) channel estimation methods~\cite{mumtaz2016mmwave} to estimate the channel. Let ${\hat{\bf{h}} }_{t}^{\rm{LS}}$ and ${\hat{\bf{h}}}_{t} ^{\rm{MMSE}}$ represent the LS and MMSE estimation results of ${\bf{h}}_{t}$, respectively, and calculate them using the following two formulas:
\begin{equation}
\label{eq: LS}
\begin{aligned}
\hat{\bf{h}}_{t} ^{\rm{LS}} = {\bf{y}}_{t},
\end{aligned}
\end{equation}
\begin{equation}
\label{eq: MMSE}
\begin{aligned}
\hat{\bf{h}}_{t} ^{\rm{MMSE} }= \mathbb{E}\left [ {\bf{h}}_{t}| {\bf{y}}_{t} \right ]  & = {\bf{\Sigma}}_{{\bf{h}}_t} \left ( {\bf{\Sigma}}_{{\bf{h}}_t}+\frac{1}{{{\mathsf{SNR}}}} {\bf{I}}_{N_{\rm{BS}}}  \right ) ^{-1} {\bf{y}}_{t}, \\
\end{aligned}
\end{equation}
where ${\bf{\Sigma}}_{{\bf{h}}_t} = \mathbb{E}\left \{ {\bf{h}}_{t}{\bf{h}}_{t}^{{\mathsf{H}}}  \right \} $ is the prior covariance matrix of channel.

\section{Problem Formulation}\label{sec: Problem Formulation}
In this section, the channel aging issue is illustrated, and the channel prediction problem is formulated to alleviate the channel aging.

\begin{figure}[!t]
	\centering
    \setlength{\abovecaptionskip}{0.cm}
	\includegraphics[width=1\linewidth]
    {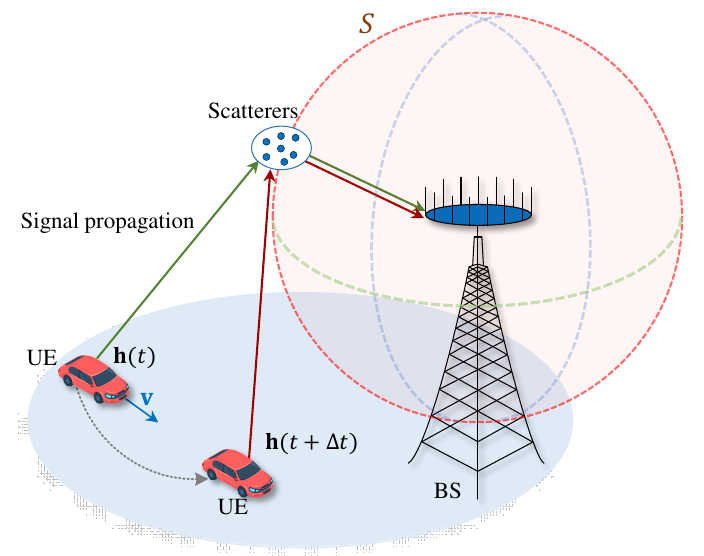}
	\caption{The XL-MIMO communication system with scatterers distributed on the spherical surface $S$ surrounding the base station. User is in motion with velocity $\bf v$.}
	\label{img: changjingtu_channel_prediction}
    \vspace{-3mm}
\end{figure}

\begin{figure}[!t]
	\centering
    \setlength{\abovecaptionskip}{0.cm}
	\includegraphics[width=1\linewidth]
    {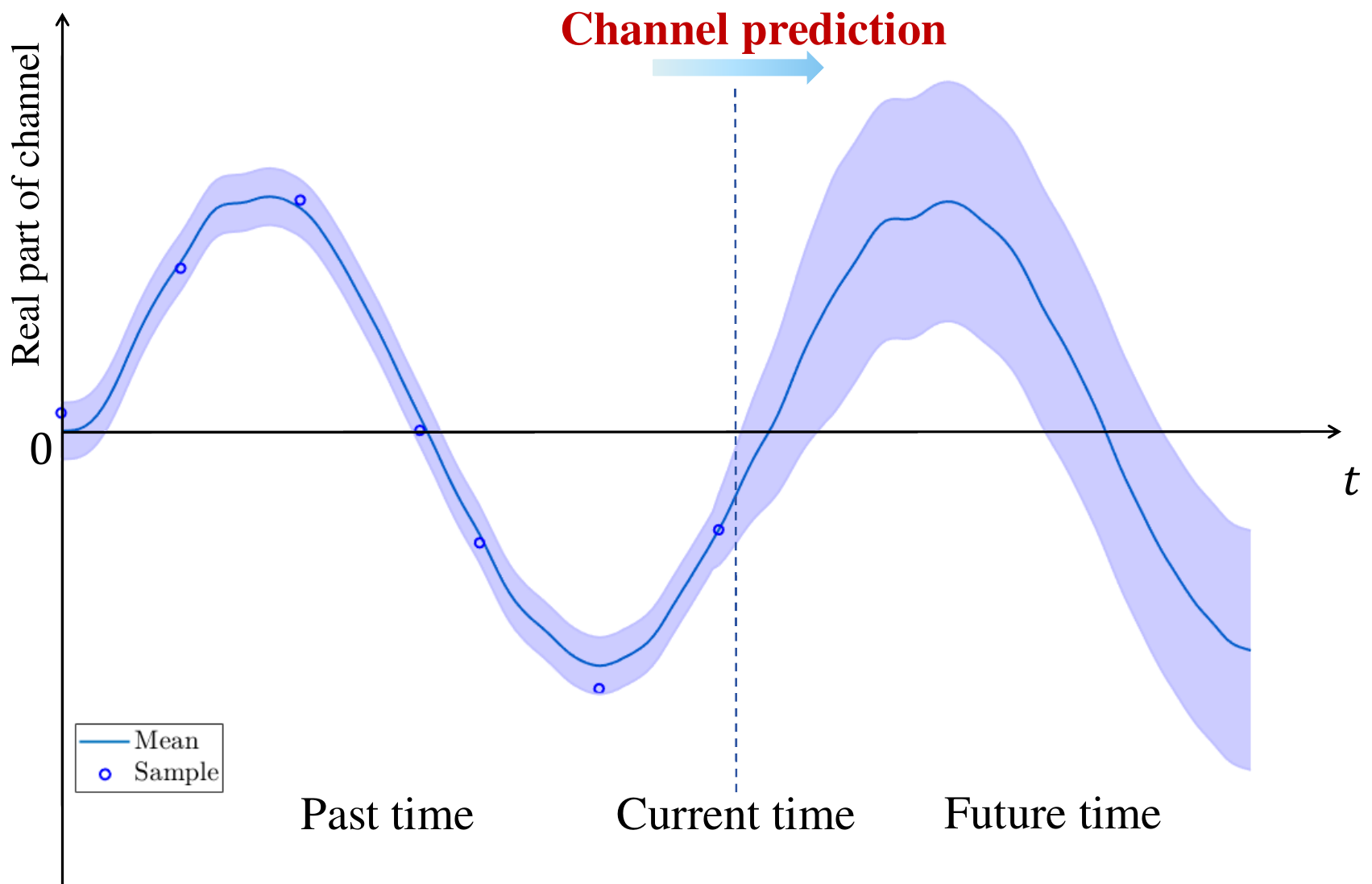}
	\caption{An illustration of channel prediction: Taking a component of a channel vector as an example, represent the variation of the channel and its uncertainty over time}
	\label{img: Illustration_channel_prediction}
    \vspace{-3mm}
\end{figure}

As shown in Fig.~\ref{img: changjingtu_channel_prediction}, in XL-MIMO communication system, the user moves at a speed $\bf v$, and the Doppler shift will cause significant differences in the channel at different times. We refer to the period of channel estimation as a frame, which contains $N_{\rm{s}}$ time slots. Channel estimation is only performed in the first slot. In mobile scenarios, because of the influence of the Doppler effect, except for the channel at the first slot, the actual time-varying channels of the follow-up slots may have significant differences from the channels obtained by the channel estimation, resulting in a decrease in the accuracy of the obtained CSI and thus affecting communication quality. Specifically, according to~\cite{cheng2008doppler}, the channel coherence time $T_{\rm{c}}$ is defined as the time during which the channel can be well regarded as time invariant, which is inversely proportional to the carrier frequency and user motion speed, i.e.,
\begin{equation}
\label{eq: coherence time}
\begin{aligned}
T_{\rm{c}}\approx \frac{c}{2fv} =\frac{\lambda }{2v} ,
\end{aligned}
\end{equation}
where $f$ is the carrier frequency, $\lambda$ is carrier wavelength and $v$ represents the user's moving speed. Channel coherence time is a rough estimate used to describe the time interval. Let $v_{\rm{r}}\le v$ represents the radial velocity relative to the BS. The calculation of Doppler shift $f_{\rm{d}}$ is 
\begin{equation}
\label{eq: Doppler shift}
\begin{aligned}
f_{\rm{d}} =\frac{v_{\rm{r}}}{\lambda} .
\end{aligned}
\end{equation}
The larger the Doppler frequency shift, the shorter the channel coherence time, and the more severe the channel aging. When the channel coherence time is shorter than the channel estimation period, using the channel estimation result of the first time slot for subsequent time slots will result in performance loss.
The variations of channel and its uncertainty due to imperfection of channel measurements over time are shown in Fig.~\ref{img: Illustration_channel_prediction}. The solid curve represents the real part of a channel vector component, and the shadow area represents its uncertainty region. It can be observed that the channel uncertainty significantly increases at future time moments.

To solve the problem of severe channel aging mentioned above, some channel prediction methods have been proposed. The channel prediction problem is to obtain future channels through past channels. The existing channel prediction methods are typically based on sequential prediction. Specifically, it is to use the channels from frame 1 to frame $T$ to predict the channel at frame $T+1$, and then use the channel at frame $T+1$ as known information to predict the channel at frame $T+2$ from frame 2 to frame $T+1$, and so on. However, due to errors in the channel prediction results at frame $T+1$, using it as a known channel to predict subsequent channels will bring errors to the subsequent predicted channels, which is the problem of error propagation.

In order to avoid performance loss caused by error propagation, unlike existing sequential channel prediction methods, we formulate the channel prediction problem in parallel form. That is, using the channel estimation results of the past $L$ channels to predict the future channel of the next $F$ channels. It should be noted that the channels of future $F$ channels are predicted simultaneously. Considering the characteristics of the GRF channel, achieving accurate channel prediction requires an appropriate autocorrelation function, i.e., the kernel. We can then predict the future channel through inference based on this kernel. The appropriate kernel form will be discussed in the next section. Let ${\bm{\omega } }\in{\bm{\Omega }}$ denote model parameters of the kernel, and ${\bm{\Omega }}$ is the set of model parameters. ${\bf{y } }=({\bf{y } }_{1}^{{\mathsf{T}}},{\bf{y } }_{2}^{{\mathsf{T}}},\dots,{\bf{y } }_{L}^{{\mathsf{T}}})^{{{\mathsf{T}}}}\in \mathbb{C}^{N_{\rm{BS}}L\times1}$ denotes the column vector composed of the received pilot sequences in the past $L$ time frames. Let $ \mathcal{L}$ denote the set of past channel indices and $ \mathcal{F}$ denote the set of future channel indices. ${\bf{h } }_{\mathcal{L} }=({\bf{h } }_{1}^{{\mathsf{T}}},{\bf{h } }_{2}^{{\mathsf{T}}},\dots,{\bf{h } }_{L}^{{\mathsf{T}}})^{{{\mathsf{T}}}}\in \mathbb{C}^{N_{\rm{BS}}L\times1}$ denotes the column vector composed of the previous $L$ channels. ${\bf{h} }_{\mathcal{F} }=({\bf{h}}_{L+1}^{{{\mathsf{T}}}},{\bf{h}}_{L+2}^{{{\mathsf{T}}}},\dots ,{\bf{h}}_{L+F}^{{{\mathsf{T}}}})^{{{\mathsf{T}}}}\in \mathbb{C}^{N_{\rm{BS}}F\times1}$ denotes the column vector composed of $F$ future channels that need to be predicted. By using ML criterion to obtain kernel parameters, and then using MMSE criterion to predict future channels, the channel prediction problem can be formulated as
\begin{equation}
\label{eq: problem formulation}
\begin{aligned}
&{\bm{\hat{\omega}}}({\bf{y } }) = \mathop{\arg\max}\limits_{{\bm{{\omega}}}\in{\bm{\Omega }}} \left\{\ln \int p({\bf{y } }|{\bf{h} }_{\mathcal{L} })p({\bf{h } }_{\mathcal{L} }|{\bm{{\omega}}})\mathrm{d}{\bf{h} }_{\mathcal{L} } \right\},
\\&{\bf{\hat{h}} }_{\mathcal{F} } ({\bf{y } } )= \mathop{\arg\max}\limits_{{\bf{h} }_{\mathcal{F} }\in{\mathbb{C}^{{N_{\rm{BS}}}F\times 1} }} \left\{\ln p({\bf{y } }|{\bf{h} }_{\mathcal{F} })+\ln p({\bf{h }}_{\mathcal{F} }|{\bm{\hat{\omega}}}({\bf{y } }))\right\}
.
\end{aligned}
\end{equation}

 In~\eqref{eq: problem formulation}, ${\bm{\hat{\omega}}}$ is the ML estimate of ${\bm{{\omega}}}$ and ${\bf{\hat{h}}}_F$ is the MMSE estimate of ${\bf{{h}}}_F$. Due to the characteristics of the GRF channel, MMSE estimation is equivalent to maximum a posteriori (MAP) estimation. ${\bf{{h}}}_{t}$ and ${\bf{{h}}}_{t+1 }$ can be used to determine the channel of the $n$-th slot. For example, for the $t$-th frame, if $0< n\le N_{\rm{s}}/2$, then determine that the channel of time slot n is ${\bf{{h}}}_{t}$. Otherwise, it is determined as ${\bf{{h}}}_{t+1}$. In the following Section~\ref{sec: Proposed}, we need to accurately solve the problem in~\eqref{eq: problem formulation}.

\section{Proposed Spatio-Temporal Electromagnetic Kernel Learning Based Channel Prediction}\label{sec: Proposed}
In this section, we propose a parallel channel prediction scheme that simultaneously utilizes the temporal and spatial EM correlation between channels to avoid error propagation and improve the accuracy of channel prediction. Firstly, in Section IV-A, we use EMCF to model the spatio-temporal correlation between channels. Then, in Section IV-B, we introduced Gaussian process regression (GPR). Moreover, in Section IV-C, the method of kernel learning is used for channel prediction. In Section IV-D, we propose grid-based electromagnetic mixed (GEM) kernel. Finally, in Section IV-E, the proposed GEM-KL channel prediction algorithm is elaborated.

\subsection{Construction of STEM correlation function}

To fully utilize the EM physical characteristics, it is essential to consider the fundamental physical principles behind the communication processes, including electromagnetics and information theory~\cite{zhu2024electromagnetic},~\cite{wan2024near}. The integration of these two theories could advance research in electromagnetic information theory (EIT), which provides insights into wireless communication issues from the perspective of electromagnetic wave propagation~\cite{chafii2023twelve,li2023electromagnetic,wan2024electromagnetic}. We use the EIT-based channel model. Specifically, based on the channel model of Gaussian random field described in subsection~\ref{sec:Channel model}, we analyze the characteristics of EM channels and their correlation. 
Electromagnetic information can be combined with the autocorrelation function of the channel~\cite{zhu2023can}. We calculate the correlation integral of the electric field on the unit sphere $S^2$ shown in Fig.~\ref{img: changjingtu_channel_prediction} to obtain the correlation function of the time-varying channel, i.e.,
\begin{equation}
\label{eq: electromagnetic autocorrelation function}
\begin{aligned}
{\bf K}({\bf x},t;
{\bf x}',t') \propto \int_{\hat{\bm \kappa}\in S^2} ({\bf I} - \hat{\bm \kappa}\hat{\bm \kappa}^{{\mathsf{T}}})e^{{\rm i}k_0 \hat{\bm \kappa}\cdot (({\bf{x}}-{\bf{x}}')+{{\bf
{v}}}(t-t'))} \nu(\hat{\bm \kappa}) {\rm d} S ,
\end{aligned}
\end{equation}
where the integration is carried out over the surface of the unit sphere $S^2 $, $k_{0} = 2{\pi}/{\lambda}_0$ is the wavenumber. $\hat{\bm \kappa}\ $ denotes the unit radial vector, and  $\nu: S^2 \rightarrow \mathbb{R}_+$  denotes the angular power spectral density of the incident wave, with units of Watts per steradian per polarization. This function is also named as electromagnetic correlation function (EMCF). For time-varying channels, we incorporate the Doppler frequency shift into the EM correlation function by introducing the velocity vector $\bf{v}$, hence this EMCF can also be referred to as the spatio-temporal kernel function (STEM-CF). To represent the incoming direction of electromagnetic waves, we use the on Mises Fisher (vMF) distribution, i.e., $\nu(\hat{\bm \kappa})=({\zeta^2}/{(8\pi)})e^{\hat{\bm \kappa} \cdot \bm{\delta}} $. ${\bm{\delta }}\in \mathbb{C}^{3}$ is the concentration parameter, and its direction represents the direction in which the electromagnetic wave is concentrated. If the electromagnetic incidence is isotropic,  $\nu(\hat{\bm \kappa}) $ is a constant ${\zeta^2}/{(8\pi)}$. It should be noted that the larger the concentration, the stronger the channel sparsity. We can compute the closed-form expression for STEM-CF as follows:

\begin{equation}
\label{eq: electromagnetic autocorrelation function}
\begin{aligned}
{\bf{K}}_{\rm STEM} ({\bm{\rho}},{\bm{\rho}{'}}) &= \mathbb{E}\left [ {\bf{E}} ({\bm{\rho}}){\bf{E}} ({\bm{\rho}}{'})^{\mathsf{H}}\right ]
\\
&=\frac{\zeta ^{2}}{S(\left \| \bm{\delta}  \right \| )} {\bm \Sigma}(\bm{\xi}) ,
\end{aligned}
\end{equation}
where ${\bf{K}}_{\rm STEM}$ is a $3\times 3$ complex matrix, $\rm{tr}({\bf{K}}_{EMCF} ({\bm{\rho}},{\bm{\rho}{'}})) = \zeta ^2$, ${\bm{\xi}} = k_{0}{\bf{w}} = k_{0}({\bf{x}} -{\bf{x}}{'} + {\bf{v}}(t-t{'}))-{\rm i}{\bm{\delta }}\in \mathbb{C}^{3}$.  $S(\delta) = {\rm sinh}(\delta)/\delta$ is an additional normalisation factor, where $\delta = \left \| \bm{\delta}  \right \| \in \mathbb{R}_{+}$. We utilize the commonly used spherical Bessel functions $j_{n}(\xi)$ in 3D scenes to represent the correlation function ${\bm \Sigma}(\bm{\xi})$ 
\begin{equation}
\label{eq: correlation function}
\begin{aligned}
 {\bm{\Sigma(\bm{\xi})}} = \frac{1}{6} (4j_{0}(\xi) -j_{2}(\xi)){\bf{I}}_3 +\frac{1}{2} (j_{2}(\xi ) -2j_{0}(\xi )){\bm{\hat{\xi}}}{\bm{\hat{\xi}}}^{{\mathsf{T}}},
\end{aligned}
\end{equation}
where $\xi = \left | \bm{\xi}  \right | =  \sqrt{\bm{\xi}^{{\mathsf{T}}}\bm{\xi} }  $,
and ${\bm{\hat{\xi}}} = {{\bm{\xi}}}/{\xi} $ denotes the normalized ${\bm{\xi}}$. The spherical Bessel function $j_{n}(\xi)$ is expressed as
\begin{equation}
\label{eq: single-variable analytic function}
\begin{aligned}
 j_{n}(\xi)= (-\xi)^{n}{\left(\frac{1}{\xi} \frac{\mathrm{d} }{\mathrm{d}\xi} \right)}^{n}\frac{\sin \xi}{\xi} ,
\end{aligned}
\end{equation}

It is important to note that ${\bf{w}} = {\bf{x}} -{\bf{x}}{'} + {\bf{v}}(t-t{'})-{\rm i} {\bm{\delta }}/k_{0}$ contains the spatial and temporal variables, which means that the correlation function we use is capable of describing the spatial and temporal correlation in an EM-consistent way. 

\subsection{Gaussian Process Regression}
Gaussian Process Regression (GPR)~\cite{schulz2018tutorial} can obtain predictions through prior information and observation data of GRF. Specifically, for the GRF $f(x)\sim \mathcal{GRF} ({{\mu}}(x), {{k}}(x,x{'} ) )$, GPR uses observation data $y_{i} = f(x_{i}) + n_{i}$, $n_{i} \sim \mathcal{CN} (0, \sigma^{2}_{n} ), i = 1,2,\dots ,L_{N}$ to get a set of $F$-point prediction $\mathcal{F} = \left \{ f(x_{L_{N}+1}), f(x_{L_{N}+2}), \dots,f(x_{L_{N}+F_{N}}) \right \} $.  where $L_N={N_{\rm{BS}}}L$ and $F_N={N_{\rm{BS}}}F$.

The joint probability distribution of the observed and predicted joint vector ${\bf{g}} = [y_{1}, y_{2}, \dots,y_{L},f(x_{L_{N}+1}), f(x_{L_{N}+2}), \dots,f(x_{L_{N}+F_{N}})]^{{\mathsf{T}}}$ satisfies 
\begin{equation}
\label{eq: joint probability distribution}
\begin{aligned}
 {\bf{g}} \sim \mathcal{CN}\left(\begin{bmatrix}
 {\bm{\mu}}_{\mathcal{L}}\\
{\bm{\mu}}_{\mathcal{F}}
\end{bmatrix},\begin{bmatrix}
 {\bf{K}}_{\mathcal{LL}}+\sigma _{n}^{2}{\bf{I}}_{{L_{N}}}  & {\bf{K}}_{\mathcal{LF}}\\
 {\bf{K}}_{\mathcal{FL}} &{\bf{K}}_{\mathcal{FF}}
\end{bmatrix} \right),
\end{aligned}
\end{equation}
where ${\bm{\mu}}_{\mathcal{L}}=[\mu(x_{1}),\mu(x_{2}), \dots,\mu(x_{L_{N}})]^{{\mathsf{T}}}$ and ${\bm{\mu}}_{\mathcal{F}}=[\mu(x_{L_{N}+1}),\mu(x_{L_{N}+2}), \dots,\mu(x_{L_{N}+F_{N}})]^{{\mathsf{T}}}$. The $(m,n)$-th entry of ${\bf{K}}_{\mathcal{LL}}\in \mathbb{C} ^{L_{N}\times L_{N}} $ is $k(x_{m},x_{n})$, for all $m,n\in\left \{ 1,\dots ,L_{N} \right \} $. The $(m,n)$-th entry of ${\bf{K}}_{\mathcal{LF}}\in \mathbb{C} ^{L_{N}\times F_{N}} $ is $k(x_{m},x_{n})$, for all $m\in\left \{ 1,\dots ,L_{N} \right \} $ and $n\in\left \{ L_{N}+1,\dots ,L_{N}+F_{N} \right \} $.
${\bf{K}}_{\mathcal{LF}}\in \mathbb{C} ^{L_{N}\times F_{N}}$and ${\bf{K}}_{\mathcal{FL}} = {\bf{K}}_{\mathcal{LF}}^{{{\mathsf{H}}}}\in \mathbb{C} ^{F_{N}\times L_{N}}$. The $(i,j)$-th entry of ${\bf{K}}_{\mathcal{FF}}\in \mathbb{C} ^{F_{N}\times F_{N}} $ is $k(x_{i},x_{j})$, for all $i,j\in\left \{ L_{N}+1,\dots ,L_{N}+F_{N} \right \} $. We use ${\bf{K}}_{{\bf{y}}}$ to represent ${\bf{K}}_{\mathcal{LL}}+\sigma _{n}^{2}{\bf{I}}_{{L_{N}}}$. From the Gaussian posterior formula~\cite{wytock2013sparse}, we can obtain
\begin{equation}
\label{eq: Gaussian posterior formula}
\begin{aligned}
 {\bm{\mu}}_{\mathcal{F| L}}&={\bm{\mu}}_{\mathcal{F}} + {\bf{K}}_{\mathcal{LF}}^{{\mathsf{H}}}{\bf{K}}_{{\bf{y}}}^{-1}{\bf{y}},
 \\
 {\bf{K}}_{\mathcal{F| L}}&={\bf{K}}_{\mathcal{F}}-{\bf{K}}_{\mathcal{LF}}^{{\mathsf{H}}}{\bf{K}}_{{\bf{y}}}^{-1}{\bf{K}}_{\mathcal{LF}},
\end{aligned}
\end{equation}
 
\begin{figure}[!t]
	\centering
    \setlength{\abovecaptionskip}{0.cm}
	\includegraphics[width=1\linewidth]
    {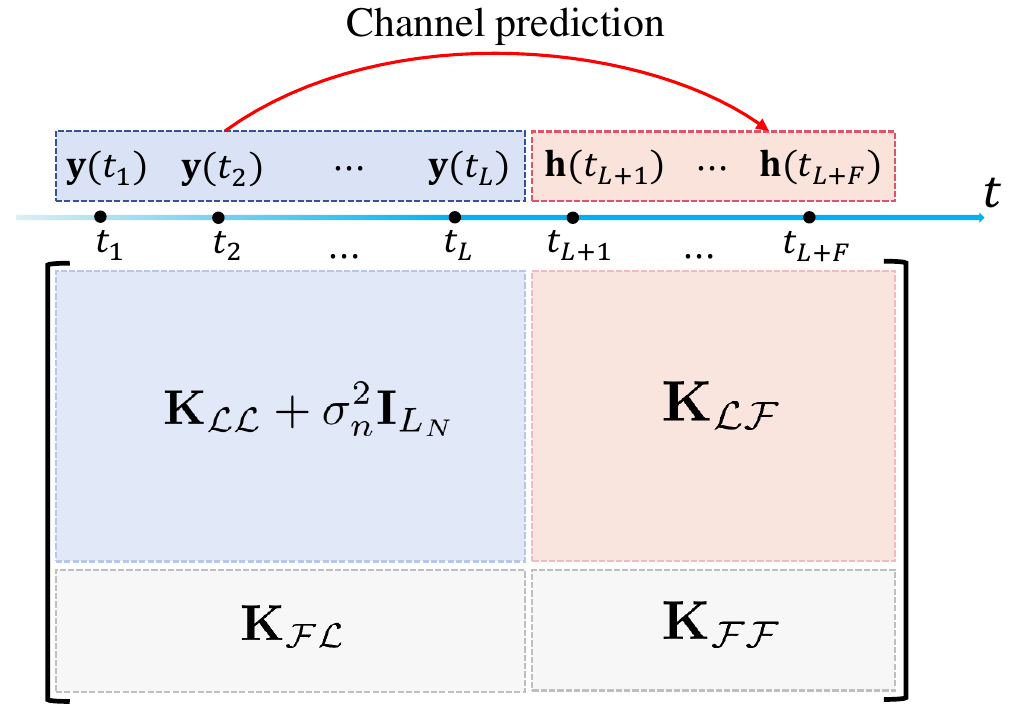}
	\caption{Gaussian process regression for time domain channel prediction.}
	\label{img: CP_K}
    \vspace{-3mm}
\end{figure}
The results of Bayesian regression are given by ${\bm{\mu}}_{\mathcal{F| L}}$ and ${\bf{K}}_{\mathcal{F| L}}$. 

As shown in Fig.~\ref{img: CP_K}, GPR channel prediction utilizes the autocovariance matrix (blue part) of the channels at past frames and the cross covariance matrix (red part) of the channels at past and future time frames to achieve parallel prediction of channels for multiple future time frames.  Since the prior distribution is a complex Gaussian distribution, the GPR predictor is consistent with the maximum a posteriori (MAP) predictor. Due to its Bayesian optimality and the degree of freedom in adjusting its kernel function parameters, it can be used for various estimation and prediction problems. Furthermore, The advantage of adjustable kernels makes GPR more widely used. The kernel adjustment measure will be introduced in subsection~\ref{sub: Kernel Learning}.

\subsection{Kernel Learning}\label{sub: Kernel Learning}
The kernel function $k(x, x')$ implicitly encodes the prior information of the Gaussian random field $f(x)$. This feature allows for more parameter configurations, thereby enhancing the model's ability to be adjusted. Choosing appropriate kernel parameters is an important step in constructing an effective regression model, which affects the accuracy of the kernel function in reconstructing Gaussian processes. The parameters that need to be adjusted in this process are usually referred to as hyperparameters. Assuming that the hyperparameters $\bm{\omega}\in \Omega \subset \mathbb{R}^{N_{{\bm{\omega}}}} $ of the adjustable kernel $k(x; x' | {\bm{\omega}})$ is also tunable. The process of finding the optimal hyperparameters for STEM kernel is called kernel learning. 

It is necessary to specify a criterion for evaluating whether hyperparameters are appropriate. The maximum likelihood (ML) criterion is a commonly used method, which can be expressed as
\begin{equation}
\label{eq: ML}
\begin{aligned}
 {\bm{\hat{\omega}}}_{\rm{ML}} =  \mathop{\arg\max}\limits_{{\bm \omega}\in \Omega } \ln{p(\bf{y}|{\bm{\omega}})},
\end{aligned}
\end{equation}
 where the probability density function (PDF) of observing $\bf{y}$ under the condition of parameter $\bm{\omega}$ is expressed as
\begin{equation}
\label{eq: probability density}
\begin{aligned}
p({\bf{y}}|{\bm{\omega}}) =  \frac{1}{\pi ^{L_{N}+F_{N}} {\rm{det}}\bf{K_{y}} } {\rm{exp}}(-{\bf{y}}^{{\mathsf{H}}}{{\bf K}_{\bf y}^{-1}} {\bf y}),
\end{aligned}
\end{equation}
The kernel ${\bf{K_{y}}} = {\bf{K_{y}}}({\bm{\omega}})$ is a function of hyperparameter ${\bm{\omega}}$. Function $l({\bm{\omega}}|{\bf{y}}) = \ln{p(\bf{y}|{\bm{\omega}})} = -\ln{\rm{det}}{\bf{K_{y}}}-(L_{N}+F_{N})\ln\pi-{\bf{y}}^{{\mathsf{H}}}{{\bf K}_{\bf y}^{-1}} {\bf y}$ is the log-likelihood function. In order to obtain the maximum likelihood estimator of the hyperparameter ${\bm{\omega}}$, methods such as gradient descent, conjugate gradient descent, and Newton iteration can be used. All of these methods require the derivative of the log-likelihood function with respect to ${\bm{\omega}}$. The calculation result of this derivative is 
\begin{equation}
\label{eq: derivative of the logarithmic likelihood function}
\begin{aligned}
\frac{\partial l({\bm{\omega}}|{\bf{y}})}{\partial \omega _{i} } = \frac{\partial }{\partial \omega _{i}} (-\ln{\rm{det}}{\bf{K_{y}}}-{\bf{y}}^{{\mathsf{H}}}{{\bf K}_{\bf y}^{-1}} {\bf y} ) \\ 
={\rm{tr}}(({\bf{g}} {\bf{g}}^{{\mathsf{H}}}-{\bf{K}}_{\bf y}^{-1})\frac{\partial {\bf{K_{y}}}}{\partial \omega _{i}} ),
\end{aligned}
\end{equation}
where $\omega _{i}$ for $i = 1,2,\dots,N_{{\bm{\omega}}}$  represents each component of hyperparameter ${\bm{\omega}}$.  For simplicity, let ${\bf{g}} = {\bf{K}}_{\bf y}^{-1}{\bf{y}}$. When the hyperparameter components are complex numbers, we need to consider the Wirtinger derivatives $({\partial }/{\partial  \omega _{{i},{\rm{Re}}}} - {\rm{i}}{\partial }/{\partial \omega _{{i},{\rm{Im}}}} )/2$. Since $l({\bm{\omega}}|{\bf{y}})$ is analytical for ${\bf{K_{y}}}$, the derivative formula ~\eqref{eq: derivative of the logarithmic likelihood function} remains unchanged.

For STEM kernel capable of predicting four-dimensional spatio-temporal channels, hyperparameters include concentration parameter $\bm{\delta }$, relative motion velocity $\bf{v }$ between transmitter and receiver, and channel energy $\zeta _{\bm{{\bf{h }} }}^2$. The Wirtinger derivatives of ${\bf{K}}_{\rm{STEM}}$ w.r.t. ${\bm{\delta}}(m)$, ${\bf{v}}(m)$ and $\zeta _{\bm{{\bf{h }} }}^2$ are respectively expressed as
\begin{equation}
\label{eq: Wirtinger derivative}
\begin{aligned}
 \frac{\partial {\bf{K}}_{\rm{STEM}}}{\partial {\bm{\delta}}(m)} &=-\frac{\zeta _{\bm{{\bf{h }} }}^2 }{S(\delta )} \left [ \frac{{S}'(\delta ){\bm{\delta}}(m) }{S(\delta)\delta}{\bm{\Sigma}} ({\bm{\xi}} )+{\rm{i}}\frac{\partial {\bm{\Sigma}} ({\bm{\xi}} )}{\partial {\bm{\xi}} (m)}   \right ],
 \\
 \frac{\partial {\bf{K}}_{\rm{STEM}}}{\partial {\bf{v}}(m)} &=\frac{\zeta _{\bm{{\bf{h }} }}^2 k_{0}(t_{p}-t_{q})}{S(\delta )} \frac{\partial {\bm{\Sigma}} ({\bm{\xi}} )}{\partial {\bm{\xi}} (m)} , 
 \\
 \frac{\partial {\bf{K}}_{\rm{STEM}}}{\partial (\zeta _{\bm{{\bf{h }} }}^2)} &=\frac{{\bm{\Sigma}} ({\bm{\xi}} ) }{S(\delta )} ,
\end{aligned}
\end{equation}
where $\delta = \left \| {\bm{\delta}} \right \| $, ${\bm{\xi}} = k_{0}{\bf{w}}$, ${\bf{w}}$ = ${\bf{x}}_{p}-{\bf{x}}_{q}+{\bf{v}}(t_{p}-t_{q})-{\rm{i}}{\bm{\delta}}/k_{0}$. The spherical Bessel functions of different orders have the following relationship
\begin{equation}
\label{eq: the property of spherical Bessel function}
\begin{aligned}
 \big(\frac{1}{\xi } \frac{\mathrm{d} }{\mathrm{d} \xi} \big)^{a}(\xi^{-b}j_{b}(\xi))=(-1)^{a}\xi^{-b-a}j_{b+a}(\xi),
\end{aligned}
\end{equation}

From the property  of spherical Bessel function~\eqref{eq: the property of spherical Bessel function}, combined with correlation function formula~\eqref{eq: correlation function}, it can be inferred that
\begin{equation}
\label{eq: Wirtinger derivative of Sigma}
\begin{aligned}
 \frac{\partial {\bm{\Sigma}} ({\bm{\xi}} )}{\partial {\bm{\xi}} (m)}&=\frac{1}{6} (-4j_{1}(\xi )-2\xi^{-1}j_{2}(\xi )+j_{3}(\xi )){\bm{\hat{\xi}}} (m){\bf{I}}_{3}
 \\
 &+\frac{1}{2} (2j_{1}(\xi )+2\xi^{-1}j_{2}(\xi )-j_{3}(\xi )){\bm{\hat{\xi}}} (m){\bm{\hat{\xi}}}{\bm{\hat{\xi}}}^{{\mathsf{T}}}
 \\
 &+\frac{1}{2} (-2j_{0}(\xi )+j_{2}(\xi ))(\partial_{m}{\bm{\hat{\xi}}}\cdot {\bm{\hat{\xi}}}^{{\mathsf{T}}}+{\bm{\hat{\xi}}}\cdot \partial_{m}{\bm{\hat{\xi}}}^{{\mathsf{T}}} ),
\end{aligned}
\end{equation}
where $\partial_{m}=\partial/\partial{\bm{\xi}} (m)$, $\xi=\left | {\bm{{\xi}}}  \right |$ and ${\bm{\hat{\xi}}} = {\bm{{\xi}}}/\xi$. Moreover, $\partial_{m}{\bm{\hat{\xi}}} = \xi^{-1}({\bf{\hat{e}}}_{m}-({\bm{\hat{\xi}}}(m)){\bm{\hat{\xi}}})$ and ${\bf{\hat{e}}}_{m}$ denotes the unit vector which the only ``1'' is located at the $m$-th component.
By combining ~\eqref{eq: derivative of the logarithmic likelihood function} and~\eqref{eq: Wirtinger derivative}, we can obtain the real-variable derivative which is expressed as
\begin{equation}
\label{eq: real-variable derivative delta}
\begin{aligned}
 \frac{\partial l}{\partial {\bm{\delta}}_{n}(m)} =  2c_{n}\mathfrak{R} \left [ {\rm{tr}}(\frac{\partial {\bf{K}}_{\mathcal{LL},{n}}}{\partial {\bm{\delta}}_{n}(m)}({\bf{g}} {\bf{g}}^{\mathsf{H}}-{\bf{K}}_{\bf y}^{-1}) )\right ] ,
\end{aligned}
\end{equation}
and
\begin{equation}
\label{eq: real-variable derivative v}
\begin{aligned}
 \frac{\partial l}{\partial {\bf{v}}_{n}(m)} =  2c_{n}\mathfrak{R} \left [ {\rm{tr}}(\frac{\partial {\bf{K}}_{\mathcal{LL},{n}}}{\partial {\bf{v}}_{n}(m)}({\bf{g}} {\bf{g}}^{\mathsf{H}}-{\bf{K}}_{\bf y}^{-1}) )\right ] ,
\end{aligned}
\end{equation}


Through gradient-based methods such as gradient ascent, these results can be used to obtain better ${\bm{\omega}}$ according to the ML criterion.
\subsection{Proposed Grid Electromagnetic Mixed  Kernel}
 The gradient based hyperparameter optimization method may get stuck in local optima. Fortunately, the grid electromagnetic mixed kernel (GEM) proposed in this subsection can achieve more global learning results.
 
 Firstly, we analyze the objective function $l({\bm{\omega}}|{\bf{y}})$, which can be intuitively represented as a function of the kernel ${\bf{K_{y}}}$. However, $l({\bm{\omega}}|{\bf{y}})$ is not a convex/concave function of ${\bf{K_{y}}}$. Therefore, gradient-based optimization methods are difficult to find the maximum value of $l({\bm{\omega}}|{\bf{y}})$. Moreover, the kernel ${\bf{K_{y}}}$ can be expressed as a function of the hyperparameter ${\bm{\omega}}$. Unfortunately, the components ${\bm{\delta}}, {\bf{v}}$ of ${\bm{\omega}}$ are not linearly related to ${\bf{K_{y}}}$, making it difficult to directly characterize the relationship between ${\bm{\omega}}$ and $l({\bm{\omega}}|{\bf{y}})$. In order to avoid the inconvenience caused by the non-convexity/concavity of functions, the grid-based method can be used in the parameter learning of STEM kernel. We design a mixed kernel composed of sub-kernels, and each of the sub-kernel corresponds to a grid point in the parameter space. In particular, several fixed values of ${\bm{\delta}}$ and $ {\bf{v}}$ are taken as the selection values for the grid.  By introducing the idea of mixed kernel, we define $k_{\rm{GEM}}$ to be a combination of multiple sub-STEM kernels. We assume that there are $N_{k}$ subcorrelation kernels and each of them has a weight of ${{{c}}}_{n}\in\mathbb{R}_{+}$, $n = 1,2,\dots,N_{k}$. Specifically, the GEM kernel function is designed as
\begin{equation}
\label{eq: mixed kernel functions}
\begin{aligned}
 k_{\rm{GEM}}&(x_{p},t_{p};x_{q},t_{q}| {\bm{{\omega}}}) 
 \\
 &= {\bf{u}}_{p}^{{\mathsf{T}}}\big(\sum_{n=1}^{N_{k}} c_{n}{\bf{K}}_{\rm{STEM}}(x_{p},t_{p};x_{q},t_{q}| {\bm{{\omega}}}_{n}) \big){\bf{u}}_{q},
\end{aligned}
\end{equation}
where the value of each $k_{\rm{GEM}}(x_{p},t_{p};x_{q},t_{q}| {\bm{{\omega}}}_{n})$ is on the grid $({\bm{\delta}}_{n},{\bf{v}}_{n})$, where ${\bm{\delta}}_{n}\in{\bm{\Delta}} $ and ${\bf{v}}_{n}\in{\bf{V}} $. And the grid values are uniformly sampled from the two-dimensional space defined by ${\bm{\Delta}} \times{\bf{V}}  $. In ~\cite{yin2020linear}, the author proved that approximating the kernel in this gridding way is effective, and will not be repeated here. ${\bm{{\omega}}}_{n}\in  \left \{ {\bm{\delta}}_{n},{\bf{v}}_{n} ,c_{n}\right \}_{n=1}^{N_{k}}\subset {\bm{{\Omega}}}$ is the collection of all the hyperparameters ${\bm{{\omega}}}_{n}\in{\bm{{\Omega}}}$. Correspondingly, the components of the mixed correlation kernel matrix can be represented as
\begin{equation}
\label{eq: mixed correlation kernel matrix}
\begin{aligned}
 ({\bf{K}}_{{\mathcal{LL}},{\rm{Mix}}})_{p,q} 
 = k_{\rm{GEM}}(x_{p},t_{p};x_{q},t_{q}| {\bm{{\omega}}}),
\end{aligned}
\end{equation}

The weight $c_{n}$ is linearly related to the kernel $k_{\rm{STEM}}(x_{p},t_{p};x_{q},t_{q}| {\bm{{\omega}}}_{n})$ in the objective function $l({\bm{\omega}}|{\bf{y}})$, so optimizing the weights $\left \{ c_{n}\right \} _{n=1}^{N_{k}}$ corresponding to different ${\bm{\delta}}_{n}$ and ${\bf{v}}_{n}$ is sufficient to obtain the optimal hyperparameters on the grid. It should be noted that all $\left \{ c_{n}\right \} _{n=1}^{N_{k}}$ are non negative, and we fix ${\textstyle \sum_{n=1}^{N_{k}}}c_{n}$ to 1.

The mixed and grid based kernel is able to improve the fitting ability of Gaussian random fields defined by STEM functions to channel observation data. In theory, a mixed kernel composed of a finite number of sub correlation functions can represent the angular power spectrum of any incident electromagnetic field. The ML problem is simplified as
\begin{equation}
\label{eq: ML_c}
\begin{aligned}
{\bf{\hat{c}}}_{\rm{ML}} =  \mathop{\arg\max}\limits_{{\bf{{c}}}} \ln{p(\bf{y}|{\bf{{c}}})},
\end{aligned}
\end{equation}
The log likelihood function is 
\begin{equation}
\label{eq: mixed log likelihood function}
\begin{aligned}
 l(\left \{ c_{n}\right \} _{n=1}^{N_{k}},\zeta ^{2}|{\bf{y}}) = &\ln{p({\bf{y}}|\left \{ c_{n}\right \} _{n=1}^{N_{k}})} 
 \\
= &-\ln{\rm{det}}{{\bf{K}}_{{{\bf{y}}},{\rm{Mix}}}}-{\bf{y}}^{{\mathsf{H}}}{{\bf{K}}^{-1}_{{{\bf{y}}},{\rm{Mix}}}}{\bf{y}}
 \\
 &+ \rm{const},
\end{aligned}
\end{equation}
where ${{\bf{K}}_{{{\bf{y}}},{\rm{Mix}}}} = {\bf{K}}_{\mathcal{LL},{\rm{Mix}}} + \sigma_{\bf{h}}^{2}{\bf{I}}_{L_{N}} = {\textstyle \sum_{n=1}^{N_{k}}}{{{c}}}_{n}{\bf{K}}_{\mathcal{LL},{n}}+ \sigma_{\bf{h}}^{2}{\bf{I}}_{L}$. $l(\left \{ {\bm{\delta}}_{n},{\bf{v}}_{n} ,c_{n}\right \} _{n=1}^{N_{k}},\zeta ^{2}|{\bf{y}})$ is the objective function. It is the fuction of ${{\bf{K}}_{{{\bf{y}}},{\rm{Mix}}}}$. Let $l_{r}(\left \{ c_{n}\right \} _{n=1}^{N_{k}},\zeta ^{2}|{\bf{y}}) = \ln{\rm{det}}{{\bf{K}}_{{{\bf{y}}},{\rm{Mix}}}}+{\bf{y}}^{{\mathsf{H}}}{{\bf{K}}^{-1}_{{{\bf{y}}},{\rm{Mix}}}}{\bf{y}}$, we transform ML problems into finding the minimum value of the objective function to eliminate negative signs.
\begin{equation}
\label{eq: ML_c_min}
\begin{aligned}
 {\bf{\hat{c}}}_{\rm{ML}} =  \mathop{\arg\min}\limits_{{\bf{{c}}}} (\ln{\rm{det}}{{\bf{K}}_{{{\bf{y}}},{\rm{Mix}}}}+{\bf{y}}^{{\mathsf{H}}}{{\bf{K}}^{-1}_{{{\bf{y}}},{\rm{Mix}}}}{\bf{y}}),
\end{aligned}
\end{equation}
where ${\bf{y}}^{{\mathsf{H}}}{{\bf{K}}^{-1}_{{{\bf{y}}},{\rm{Mix}}}}{\bf{y}}$ is a convex function about ${{\bf{K}}_{{{\bf{y}}},{\rm{Mix}}}}$ and $\ln{\rm{det}}{{\bf{K}}_{{{\bf{y}}},{\rm{Mix}}}}$ is a concave function about ${{\bf{K}}_{{{\bf{y}}},{\rm{Mix}}}}$. The majorization-minimization (MM) algorithm~\cite{sun2016majorization} can be used to solve the optimal hyperparameters with non-convex and non-concave objective functions through an iterative scheme. Each iteration must minimize the designed surrogate function. 

In the majorization step, we use the first-order Taylor expansion to design the surrogate function, which approximates the upper bound of the concave part of the function. Linearization of $\ln{\rm{det}}{{\bf{K}}_{{{\bf{y}}},{\rm{Mix}}}}$ at ${{\bf{K}}_{{{\bf{y}}},{\rm{Mix}}}}={{\bf{K}}^{(m)}_{{{\bf{y}}},{\rm{Mix}}}}$ $({\bf{{c}}}={\bf{{c}}}^{(m)})$ yields the following inequality:
\begin{equation}
    \label{eq: inequality}
    \begin{aligned}
     l_{r}({{\bf{K}}_{{{\bf{y}}},{\rm{Mix}}}}) \le &{\bf{y}}^{{\mathsf{H}}}{{\bf{K}}^{-1}_{{{\bf{y}}},{\rm{Mix}}}}{\bf{y}} + l_{\rm{CCV}}({{\bf{K}}^{(m)}_{{{\bf{y}}},{\rm{Mix}}}})
     \\
     +& {\rm tr}\left(\nabla l_{\rm{CCV}}({{\bf{K}}^{(m)}_{{{\bf{y}}},{\rm{Mix}}}})^{{\mathsf{T}}}({{\bf{K}}_{{{\bf{y}}},{\rm{Mix}}}}-{{\bf{K}}^{(m)}_{{{\bf{y}}},{\rm{Mix}}}})\right),
    \end{aligned}
\end{equation}
where $l_{\rm{CCV}}({{\bf{K}}^{(m)}_{{{\bf{y}}},{\rm{Mix}}}}) = \ln{\rm{det}}{{\bf{K}}^{(m)}_{{{\bf{y}}},{\rm{Mix}}}}$ and ${\big(\nabla l({{\bf{K}}})\big)_{ij}}={\partial l}/{\partial {{\bf{K}}}_{ij}}$.
The Wirtinger derivative of $l$ w.r.t. ${\bf{K}}_{\mathcal{LL},{n}}$ is given by the following formula
\begin{equation}
\label{eq: the Wirtinger derivative of l w.r.t. K}
\begin{aligned}
 \frac{\partial l}{\partial {\bf{K}}_{\mathcal{LL},{n}}} =  ({\bf{g}} {\bf{g}}^{\mathsf{H}}-{{\bf{K}}^{-1}_{{{\bf{y}}},{\rm{Mix}}}})^{\ast },
\end{aligned}
\end{equation}
where ${\bf{g}}={\bf{K}}_{{{\bf{y}}},{\rm{Mix}}}^{-1}{\bf{y}}$. The real-variable derivative of the objective function $l$ with respect to $c_{n}$ is expressed as
\begin{equation}
\label{eq: real-variable derivative c}
\begin{aligned}
 \frac{\partial l}{\partial c_n} =  2\mathfrak{R} \left [ {\rm{tr}}({\bf{K}}_{\mathcal{LL},{n}}({\bm{\omega}}_{n})({\bf{g}} {\bf{g}}^{\mathsf{H}}-{\bf{K}}_{{{\bf{y}}},{\rm{Mix}}}^{-1} )\right ] ,
\end{aligned}
\end{equation}

Using formulas ~\eqref{eq: inequality}, ~\eqref{eq: the Wirtinger derivative of l w.r.t. K} and ~\eqref{eq: real-variable derivative c}, the surrogate function $l_s$ of the MM algorithm is written as 
\begin{equation}
\label{eq: surrogate function}
\begin{aligned}
 l_{s}({\bf{{c}}}| {\bf{{c}}}^{\rm{(m)}}) = &{\bf{y}}^{{\mathsf{H}}}{{\bf{K}}^{-1}_{{{\bf{y}}},{\rm{Mix}}}}{\bf{y}} + \ln{\rm{det}}{{\bf{K}}^{(m)}_{{{\bf{y}}},{\rm{Mix}}}}
 \\
 +&2\mathfrak{R} \left \{ {\rm{tr}}\left [\big({({{\bf{K}}^{(m)}_{{{\bf{y}}},{\rm{Mix}}}})^{-1}}\big)^{{\mathsf{T}}}({{\bf{K}}_{{{\bf{y}}},{\rm{Mix}}}}-{{\bf{K}}^{(m)}_{{{\bf{y}}},{\rm{Mix}}}}) \right ] \right \},      
\end{aligned}
\end{equation}

Then, in the minimization step, the weight $\left \{ c_{n}\right \} _{n=1}^{N_{k}}$ is updated through
\begin{equation}
\label{eq: minimization step}
\begin{aligned}
 {\bf{\hat{c}}}^{\rm{(m+1)}} =  \mathop{\arg\min}\limits_{{\bf{{c}}}} (l_{s}({\bf{{c}}}| {\bf{{c}}}^{\rm{(m)}}) ),
\end{aligned}
\end{equation}
The minimization step can be solved by finding the minimum value point of the convex function $l_{s}({\bf{{c}}}| {\bf{{c}}}^{\rm{(m)}})$, which requires the real-variable derivative of the surrogate function with respect to $c_n$
\begin{equation}
\label{eq: real-variable derivative of the surrogate function with respect to c_n}
\begin{aligned}
 \frac{\partial l_{s}}{\partial c_n} =  2\mathfrak{R} \left [ {\rm{tr}}\left({\bf{K}}_{\mathcal{LL},{n}}({\bm{\omega}}_{n})\big(({{\bf{K}}^{(m)}_{{{\bf{y}}},{\rm{Mix}}}})^{-1}-{\bf{g}} {\bf{g}}^{{\mathsf{H}}}\big) \right)\right ] ,
\end{aligned}
\end{equation}
These can be used for iteratively solving the optimal weight $\left \{ c_{n}\right \} _{n=1}^{N_{k}}$ in the MM algorithm.
The sequence $\big(l_{r}({\bf{{c}}}^{(m)})\big)_{m\in\mathbb{N} }$ is non-increasing since
\begin{equation}
\label{eq: sequence non-increasing}
\begin{aligned}
 l_{r}({\bf{{c}}}^{(m+1)})\le l_{s}({\bf{{c}}}^{(m+1)}|{\bf{{c}}}^{(m)})\le l_{s}({\bf{{c}}}^{(m)}|{\bf{{c}}}^{(m)})=l_{r}({\bf{{c}}}^{(m)}),
\end{aligned}
\end{equation}

 The first term in the objective function~\eqref{eq:  mixed log likelihood function} represents model complexity, while the second term represents data fitness. Kernel learning needs to balance these two factors. The process of maximizing the objective function $l$ is capable of automatically balancing model complexity and data fitness. The GEM kernel parameter learning algorithm is summarized in {\bf{Algorithm}~\ref{alg:3}}, and in the next subsection we will summarize the overall GEM channel prediction algorithm.
\begin{algorithm}[!t] 
	\caption{Proposed GEM Kernel Parameter Learning Algorithm.} 
	\label{alg:3} 
	\begin{algorithmic}[1] 
		\REQUIRE  
		Number of sub-kernels $N_{k}$; grid hyperparameters $\left \{ {\bm{\delta}}_{1},{\bm{\delta}}_{2} ,\dots,{\bm{\delta}}_{N_{k}}\right \}$ and $\left \{ {\bf{v}}_{1},{\bf{v}}_{2} ,\dots,{\bf{v}}_{N_{k}}\right \}$; Received pilots $\left \{ y_{1},y_{2} ,\dots,y_{L_{N}}\right \}$; Noise variance $\sigma_{\bf{h}}^{2}$; Maximum iteration number $M_{\rm{iter}}$.
		\ENSURE  
        Hyperparameters learning results $\left \{ {\bm{\delta}}_{n},{\bf{v}}_{n} ,{\hat{c}_{n}}\right \}_{n=1}^{N_{k}}$; ${\hat \zeta ^{2}}$.
        \STATE Initialization: $\left \{ c^{(0)} \right \}_{n=1}^{N_{k}} $, learning rates of Armijo-Goldstein’s optimizer.
        \STATE Set $m\gets 0$.
        \STATE Let ${\bf{y}}\in \mathbb{C } ^{L_{N}\times 1}$ containing received pilots from $\left \{ y_{1},y_{2} ,\dots,y_{L_{N}}\right \}$.
		\FOR{$m = 1,2,\dots,M_{\rm{iter}}$}
        \STATE Construct the GEM kernel ${{\bf{K}}_{{{\bf{y}}},{\rm{Mix}}}}$ from hyperparameters $\left \{ {\bm{\delta}}_{n}^{(m-1)},{\bf{v}}_{n}^{(m-1)} ,{{c}_{n}}^{(m-1)}\right \}_{n=1}^{N_{k}}$ by ~\eqref{eq: mixed kernel functions} and ~\eqref{eq: mixed correlation kernel matrix}.
        \STATE ${\bf{g}}\gets {{\bf{K}}^{-1}_{{{\bf{y}}},{\rm{Mix}}}}{\bf{y}}$
		\FOR{$n=1,2,\dots, N_{k}$}
		\STATE Construct surrogate function $l_{s}({{{c}}}_{n}| {{{c}}}^{{(m)}}_{n})$ by ~\eqref{eq: surrogate function}. 
        \STATE Compute $\frac{\partial l_{s}}{\partial c_n}$ from  ~\eqref{eq: real-variable derivative of the surrogate function with respect to c_n}.
        \STATE Update ${{{c}}}^{{(m)}}_{n}$ from ~\eqref{eq: surrogate function}. by Armijo-Goldstein’s optimizer.
        \STATE Update $\left \{ c^{(m)} \right \}_{n=1}^{N_{k}}$ from ${{{c}}}^{{(m)}}_{n}$.
        \STATE Update ${{\bf{K}}_{{{\bf{y}}},{\rm{Mix}}}}$ from $\left \{ c^{(m)} \right \}_{n=1}^{N_{k}}$.
		\ENDFOR
		\ENDFOR
        \STATE ${\hat \zeta ^{2}} \gets 2 {\textstyle \sum_{\ell=1 }^{L_{N}}} \left | y_{\ell } \right |^{2} /\big(L_{N}\cdot (1+\sigma_{\bf{h}}^{2})\big)$
		\RETURN Hyperparameters learning results $\left \{ {\bm{\delta}}_{n},{\bf{v}}_{n} ,{\hat{c}_{n}}\right \}_{n=1}^{N_{k}}$, and  ${\hat \zeta ^{2}}$.
	\end{algorithmic}
\end{algorithm}

\subsection{Proposed GEM-KL Channel Prediction Algorithm}
We set the number of base station antennas to ${N_{\rm{BS}}}$, assuming that these antennas are located at $\left \{ {\bf{x}}_{n}  \right \}_{n=1}^{N_{\rm{BS}}} \subset \mathbb{R}^{3}$. We consider the spatio-temporal correlation tensor between the $m$-th polarization of antenna $a$ at time $t_i$ and the $n$-th polarization of antenna $b$ at time $t_j$. Let $p=(a,m,i) $ and $q=(b,n,j) $ , the correlation tensor cam be expressed as
\begin{equation}
\label{eq: correlation tensor}
\begin{aligned}
 {\bf{K}}_{p ,q } =  {\bf{u}}_{p}^{{\mathsf{T}}} \left [ {\bf{K}}_{\rm{STEM}} ( {\bf{x}}_{p},t_{p}; {\bf{x}}_{q},t_{q}) \right ]  {\bf{u}}_{q},
\end{aligned}
\end{equation}
where ${\bf{u}}_{p}$ represents the unit vector of antenna polarization direction. On the basis of formula~\eqref{eq: correlation tensor},  correlation matrix between several channels in different time and space can be calculated, and the specific scheme is given by {\bf{Algorithm}~\ref{alg:1}}. The proposed EIT based GPR channel predction method is summarized in {\bf{Algorithm}~\ref{alg:2}}. Specifically, the BS receives noisy observations at any spatio-temporal coordinate at past times and predict the channel at future times. In this algorithm, the unknown channels in the future or past time are modeled as Gaussian random field. We need to first use STEM-CF to calculate the autocorrelation matrix ${\bf{K}}_{{\bf{y}}}={\bf{K}}_{\mathcal{LL}}+\sigma _{n}^{2}{\bf{I}}_{{L}}$ of the channels at past times. And then calculate the correlation matrix between the past and future channels. Finally we use ~\eqref{eq: Gaussian posterior formula} to obtain the future channels. The performance of the proposed channel prediction algorithm will be evaluated in the next section.
\begin{algorithm}[!t] 
	\caption{Channels Correlation Matrix Design.} 
	\label{alg:1} 
	\begin{algorithmic}[1] 
		\REQUIRE  
		GEM hyperparameters ${\bm{\omega}}=\left \{ {\bm{\delta}}_{n},{\bf{v}}_{n} ,{\hat{c}_{n}}\right \}_{n=1}^{N_{k}}\in \Omega$, channel indices $p\in\mathcal{P} $, $q\in\mathcal{Q} $, $p_{\rm min}$,  $p_{\rm max}$, $q_{\rm min}$, $q_{\rm max}$.
		\ENSURE  
        The correlation matrix between the channels in set $\mathcal{P}$ and the channels in set $\mathcal{Q}$: ${\bf{K}}_{\mathcal{PQ}}$.
		\STATE Let ${\bf{K}}_{\mathcal{PQ}}\in {\mathbb{C}}^\mathcal{\left | P \right | \times \left | Q \right | }$, $p=p_{\rm min}$, $q=q_{\rm min}$.
		\FOR{$p=p_{\rm min},p_{\rm min}+1,\dots, p_{\rm max}$}
		\FOR{$q=q_{\rm min},q_{\rm min}+1,\dots, q_{\rm max}$}
		\STATE Calculate the STEM function: ${\bf{K}}_{{pq}}\gets  {\bf{u}}_{p}^{{\mathsf{T}}}{\bf{K}}_{\rm{STEM}}( {\bf{x}}_{p},t_{p}; {\bf{x}}_{q},t_{q}|{\bm{\omega}}){\bf{u}}_{q}$ according to~\eqref{eq: electromagnetic autocorrelation function}. 
		\ENDFOR
		\ENDFOR
		\RETURN The correlation matrix ${\bf{K}}_{\mathcal{PQ}}$.
	\end{algorithmic}
\end{algorithm}
\begin{algorithm}[!t] 
	\caption{Proposed EIT-GEM Channel Predictor.} 
	\label{alg:2} 
	\begin{algorithmic}[1] 
		\REQUIRE  
		Past channel indices $l\in \mathcal{L}$; future channel indices $f\in \mathcal{F}$; Received pilots $y_{{l}},l\in \mathcal{L}$; EMCF hyperparameters ${\bm{\omega}}$;  Noise variance $\sigma _{n}^{2}$.
		\ENSURE  
        Channel prediction result ${\bf{\hat{h}}}_{\mathcal{F}}$.
        \STATE Obtain GEM hyperparameters $\left \{ {\bm{\delta}}_{n},{\bf{v}}_{n} ,{\hat{c}_{n}}\right \}_{n=1}^{N_{k}}$ according to {\bf{Algorithm}~\ref{alg:3}}.
		\STATE Compute the correlation matrix of past channels ${\bf{K}}_{\mathcal{LL}}$ and the correlation matrix between the past channels and the future channels ${\bf{K}}_{\mathcal{FL}}$ according to {\bf{Algorithm}~\ref{alg:1}}.
		\STATE ${\bf{K}}_{\bf{y}} = {\bf{K}}_{\mathcal{LL}} + \sigma _{n}^{2}{\bf{I}}_{\mathcal{\left | L \right | \times \left | L \right |} }$.
		\STATE ${\bf{g}} \gets {\bf{K}}_{\bf{y}}^{-1}{\bf{y}}$.
		\STATE Reconstruct the predicted futrue channels ${\bf{\hat{h}}}_{\mathcal{F}} \gets {\bf{K}}_{\mathcal{FL}}{\bf{g}}$ according to ~\eqref{eq: Gaussian posterior formula}.
		\RETURN The prediction result of vectorized future channels ${\bf{\hat{h}}}_{\mathcal{F}}$.
	\end{algorithmic}
\end{algorithm}

\section{Simulation Results}\label{sec: sim}
The simulation results of STEM-KL and GEM-KL channel predictor are provided in this section. We evaluate the statistical learning performance of the proposed GEM covariance predictor by comparing it to the traditional methods.
\subsection{Simulation Setup}
In the following channel prediction simulation, in order to ensure the realness of the channel, we evaluated the performance of various prediction algorithms using standard 3GPP TR 38.901 CDL model and multipath near-field Saleh--Valenzuela (SV) channel model~\cite{lu2023near}, respectively. 
The near-field multipath SV channel between the user and the BS at time $t$ can be represented as
\begin{equation}
\label{eq: the channel from the user to the BS}
\begin{aligned}
{\bf{h}}_{t} = \sum_{l=1}^{L} a_l e^{{\rm i}k_{0}r_l}{\bf{b}}(\phi_l(t),r_l(t)),
\end{aligned}
\end{equation}
where $L$ is the number of propagation paths;  $a_l$, $\phi$ and $r$ are the complex gain,  spatio angle, and the distance of the $l$-th path, respectively. 
 The uniform linear arrays (ULAs) is taken into consideration. Then, the near field array steering vector ${\bf{a}}(\phi,r)$ could be expressed by
\begin{equation}
\label{eq: near field response}
\begin{aligned}
{\bf{b}}( \phi,r) = \frac{1}{\sqrt{N}}\left[e^{-{\rm i}k_0(r^{(-\widetilde{N})}-r)},\cdots,e^{-{\rm i}k_0(r^{(\widetilde{N})}-r)}\right]^{\mathsf H},
\end{aligned}
\end{equation}
where $k_0=2\pi/\lambda$ denotes the wavenumber, $r$ denotes the distance between the scatterer (or UE) and the center of the array, and $r^{(n)}$ denotes the distance between the scatterer (or UE) and the $n$-th antenna element. We assume $N_{\rm BS}$  to be  and the maximum index is $\widetilde{N} = N_{\rm BS}/{2}$. Based on the spherical-wave propagation model, the distance $r_l^{(n)}$ of the $l$-th path can be denoted as
\begin{equation}
\label{eq: near field distance term}
\begin{aligned}
r_{l}^{(n)} &= \sqrt{r_l^2+n^2d^2-2ndr_l\sin\phi_l}, \\
\end{aligned}
\end{equation}
for the multipath near-field SV channel model, it contains $L = 10$ NLoS path components, and the Rician factor is $10$ dB. The sampled angles of departure follow the uniform distribution $\mathcal{U} \big(-\pi /3,\pi /3\big)$. Meanwhile, the distance between the base station center and the user is within the range of 5 to 30 meters. 

For the CDL channel model, the standard CDL-A delay profile is adopted.

The system parameter settings are as follows: The 256-element array is considered in simulations. The center of the antenna array is located at $(0,0,0)$, ULA is located on the $x$-axis, and the user moves in the $xoz$ plane. And the carrier frequency is set to $f_{c} = 3.5$ GHz. This means that the carrier wavelength is $0.0857\,{\rm m}$. And the array is half-wavelength space. We set period of transmitting pilot signals to 0.625 ms. The unit vector of antenna polarization direction is ${\bf{u}}=(0,1,0)^{{\mathsf{T}}}$.

{\it{Baseline algorithms.}}
The no prediction NMSE is obtained by comparing the current channel with the future channel. The AR predictor is given by the autoregressive modeling~\cite{yusuf2021autoregressive}. The PVEC predictor is reproduced from the prony vector prediction method proposed in~\cite{yin2020addressing}. 

All channel prediction algorithms are evaluated using normalized mean square error (NMSE) performance, which is defined in ~\eqref{eq: NMSE}.
\begin{equation}
\label{eq: NMSE}
\begin{aligned}
 {\rm{NMSE}} = \mathbb{E} \left[ \frac{\left \| {\bf{\hat{h}}}_{t}-{\bf{{h}}}_{t} \right \| ^{2}}{\left \| {\bf{{h}}}_{t} \right \| ^{2}}  \right] ,
\end{aligned}
\end{equation}

\subsection{Simulation Results on Multipath Near-field SV Channel }
\begin{figure}[!t] 
	\centering
	\setlength{\abovecaptionskip}{0.cm}
	\includegraphics[scale=0.755]{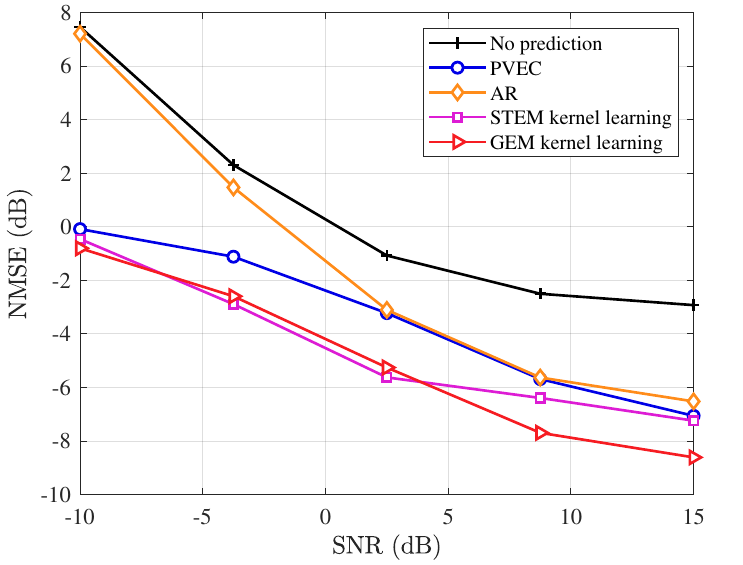}
	\caption{The NMSE performance versus SNR in multipath near-field SV channel model at $v=36\,{\rm km/h}$.}
	\label{img: near_v_10_SNR_NMSE1}
    \vspace{-3mm}
\end{figure}

\begin{figure}[!t] 
	\centering
	\setlength{\abovecaptionskip}{0.cm}
	\includegraphics[width=0.99\linewidth]{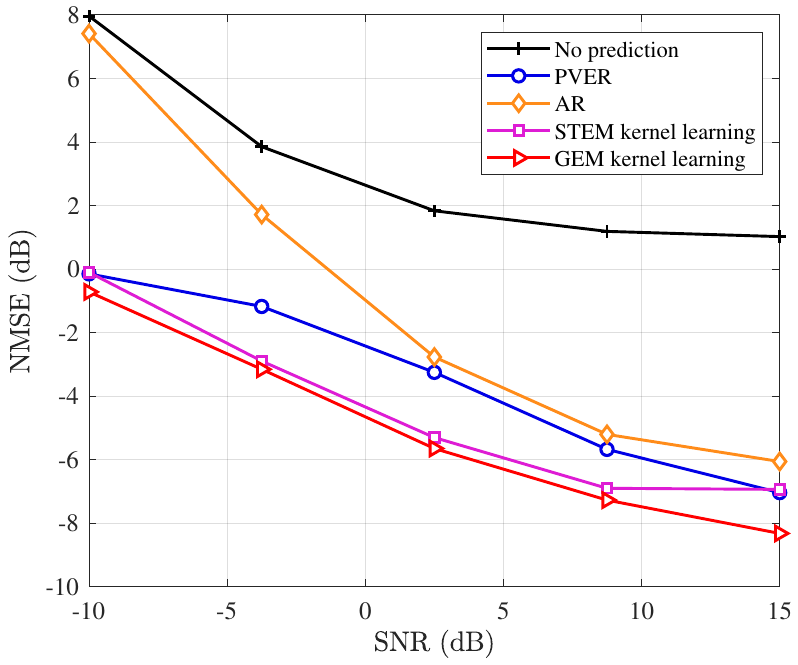}
	\caption{The NMSE performance versus SNR in multipath near-field SV channel model at $v=72\,{\rm km/h}$.}
	\label{img: near_v_20_SNR_NMSE1}
    \vspace{-3mm}
\end{figure}

In this subsection, we compare the performance of traditional channel prediction schemes with the proposed EIT-based channel prediction scheme using the multipath near-field SV channel model. First, we compare the NMSE performance of different methods for predicting the channel at the next moment as a function of SNR, the simulation results of are shown in Fig.~\ref{img: near_v_10_SNR_NMSE1} and Fig.~\ref{img: near_v_20_SNR_NMSE1}. We set the user speeds to $36\,{\rm km/h}$ in Fig.~\ref{img: near_v_10_SNR_NMSE1} and $72\,{\rm km/h}$ in Fig.~\ref{img: near_v_20_SNR_NMSE1}. From Fig.~\ref{img: near_v_10_SNR_NMSE1} and Fig.~\ref{img: near_v_20_SNR_NMSE1}, it can be seen that the channel prediction method based on kernel learning proposed in this article is significantly better than traditional methods across an SNR range of $-10\sim 15$ dB, especially in low signal-to-noise ratio situations. Among them, the grid based electromagnetic (GEM) kernel learning method can achieve the lowest NMSE. For example, when ${\mathsf{SNR}}=2.5\,{\rm dB}$, compared with the AR channel prediction method, the GEM kernel learning channel prediction scheme can achieve NMSE performance gains of $2.1\,{\rm dB}$ and $2.9\,{\rm dB}$ for the next channel prediction at $v=36\,{\rm km/h}$ and $v=72\,{\rm km/h}$, where scalar $v=\|{\bf v}\|$ is the user's moving speed. 


Compared to other channel prediction algorithms, the reason why the electromagnetic kernel based scheme performs better is mainly because the electromagnetic prior information is successfully embedded into the STEM-CF covariance model used, so the prior information provided by the electromagnetic kernel is more accurate, thus enabling more accurate channel prediction. The performance of EM channel prediction methods with kernel learning is superior to all baseline methods and non learning EM channel prediction methods, because kernel learning based channel prediction methods can obtain more accurate model hyperparameters through learning, which enables EM kernels to better fit the direct covariance function of the channel and provide more accurate prior information. The GEM kernel learning scheme outperforms all rivals mainly because it solves the problem of EM kernel learning methods falling into local optima during hyperparameter learning. It transforms the optimization of concentration $\bm\delta$ and user speed $\bf v$ into the optimization of weights for kernels composed of different $\bm\delta$ and $\bf v$. The mixed kernel approach can better adapt to multipath channels and has a strong ability to match electromagnetic correlation patterns in received pilots in the past times. Therefore, the prior information of GEM is more accurate, resulting in more stable and accurate performance.

\begin{figure}[!t]
	\centering
	\setlength{\abovecaptionskip}{0.cm}
	\includegraphics[width=1\linewidth]{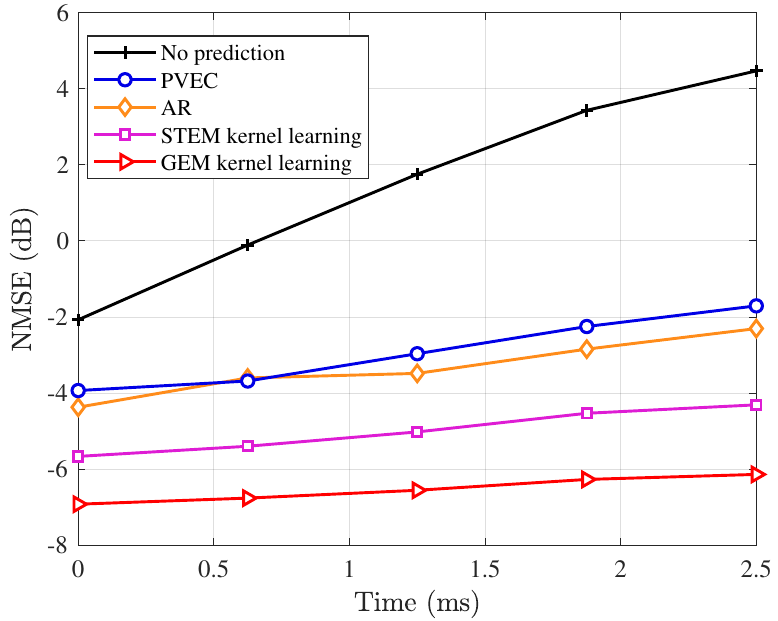}
	\caption{The NMSE performance versus time in multipath near field SV channel model at $36\,{\rm km/h}$.}
	\label{img: near_v_10_time_NMSE}
    \vspace{-3mm}
\end{figure}

\begin{figure}[!t]
	\centering
	\setlength{\abovecaptionskip}{0.cm}
	\includegraphics[width=1\linewidth]{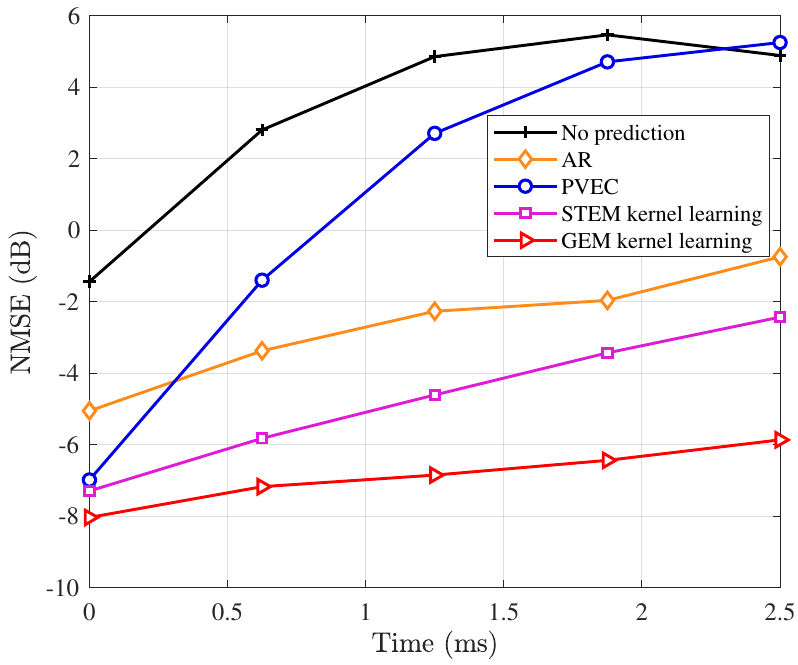}
	\caption{The NMSE performance versus time in multipath near field SV channel model at $72\,{\rm km/h}$.}
	\label{img: near_v_20_time_NMSE1}
    \vspace{-3mm}
\end{figure}

In order to investigate the performance changes of the algorithm over times, we use the channels of the past two time periods to predict the channels of the next five time periods, i.e. $L=2$ and $F=5$. We observe the NMSE performance of the channels predicted by different schemes at different time periods through simulation. Under the condition of ${\mathsf{SNR}}=5\,{\rm dB}$ and $v=36\,{\rm km/h}$, the NMSE against time  is plotted in Fig.~\ref{img: near_v_10_time_NMSE}. When ${\mathsf{SNR}}=5\,{\rm dB}$ and $v=72\,{\rm km/h}$, the corresponding performance comparison simulation results are shown in Fig.~\ref{img: near_v_20_time_NMSE1}. The different simulation points on the time scale represent the NMSE of channel prediction for different future time periods. From the simulation results, we can observe that when predicting several future channels using a small number of past time channels, the NMSE performance of the EM kernel based channel prediction algorithm is far superior to the baseline algorithm. Among them, the proposed learning based GEM kernel learning method performing the best. Taking the prediction of the fifth channel in the future as an example, the GEM kernel learning scheme proposed in this paper improves the NMSE performance by $3.8\,{\rm dB}$ and $5.1\,{\rm dB}$ respectively compared to the AR channel prediction method in the scenarios of $36\,{\rm km/h}$ and $72\,{\rm km/h}$.


It is worth noting that when predicting multiple future time channels, as time increases, the NMSE growth of the STEM kernel based methods are slower compared to the baseline method, indicating more stable performance. This is because the channel prediction methods based on STEM kernel can achieve parallel prediction of channels at multiple time points, avoiding the propagation of prediction errors.

By summarizing the simulation results of Fig.~\ref{img: near_v_10_SNR_NMSE1}$\sim$\ref{img: near_v_20_time_NMSE1}, it can be concluded that the proposed GEM kernel learning method can achieve higher accuracy in predicting future channels. In addition, this scheme can effectively alleviate the negative impact of user mobility on wireless 
 communication.

\subsection{Simulation Results on the CDL Channel }
Next, we consider the CDL-A channel model generated by Matlab 5G Toolbox.

\begin{figure}[!t]
	\centering
	\setlength{\abovecaptionskip}{0.cm}
	\includegraphics[width=1\linewidth]{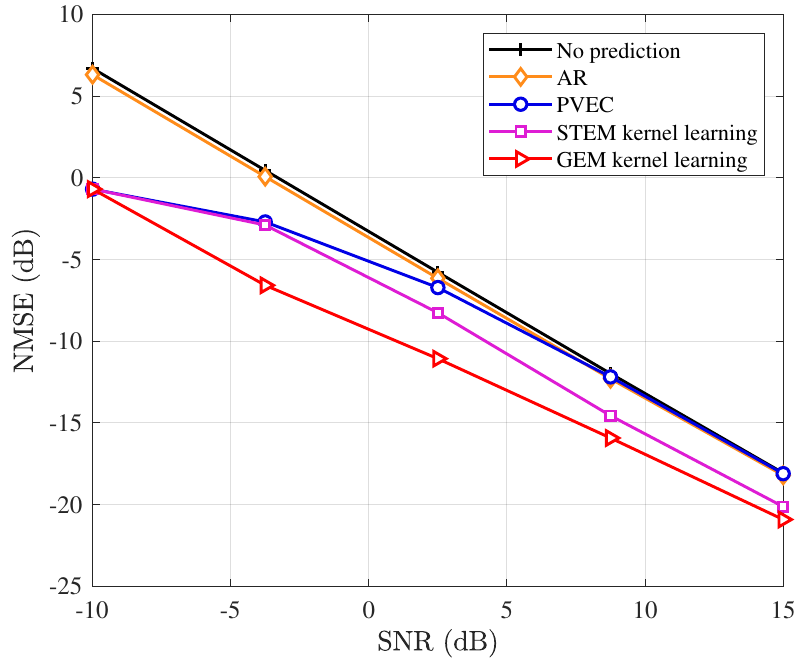}
	\caption{Comparison of the NMSE performance versus SNR between the proposed EM kernel learning channel prediction method and traditional channel prediction schemes in CDL channel scenarios at the maximum Doppler velocity of $36\,{\rm km/h}$.}
	\label{img: CDL_SNR_NMSE}
    \vspace{-3mm}
\end{figure}

\begin{figure}[!t]
	\centering
	\setlength{\abovecaptionskip}{0.cm}
	\includegraphics[width=1\linewidth]{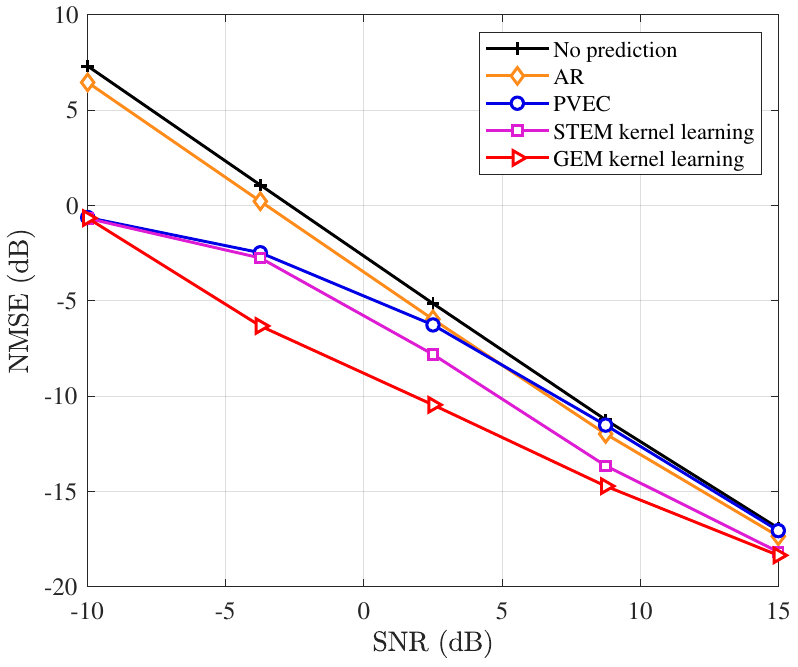}
	\caption{Comparison of the NMSE performance versus SNR between the proposed EM kernel learning channel prediction method and traditional channel prediction schemes in CDL channel scenario at the maximum Doppler velocity of $72\,{\rm km/h}$.}
	\label{img: CDL_SNR_NMSE_v_20}
    \vspace{-3mm}
\end{figure}

The trends of NMSE versus SNR for different channel prediction schemes are plotted in Fig.~\ref{img: CDL_SNR_NMSE} and Fig.~\ref{img: CDL_SNR_NMSE_v_20}. From the simulation results, it can be observed that several STEM based channel prediction schemes perform better than non prediction scheme, AR scheme, and PVEC scheme in CDL channel scenarios with maximum Doppler velocities of $36\,{\rm km/h}$ and $72\,{\rm km/h}$ (i.e. Doppler shifts of approximately $116.69\,{\rm Hz}$ and $233.37\,{\rm Hz}$), respectively. Among them, the GEM-KL scheme can achieve the best performance in both scenarios.

For example, at ${\mathsf{SNR}}=2.5\,{\rm dB}$, compared to the AR scheme, the GEM kernel learning scheme can achieve NMSE performance gains of approximately $5\,{\rm dB}$ and $4.5\,{\rm dB}$ at $36\,{\rm km/h}$ and $72\,{\rm km/h}$, respectively.

\begin{figure}[!t]
	\centering
	\setlength{\abovecaptionskip}{0.cm}
	\includegraphics[width=1\linewidth]{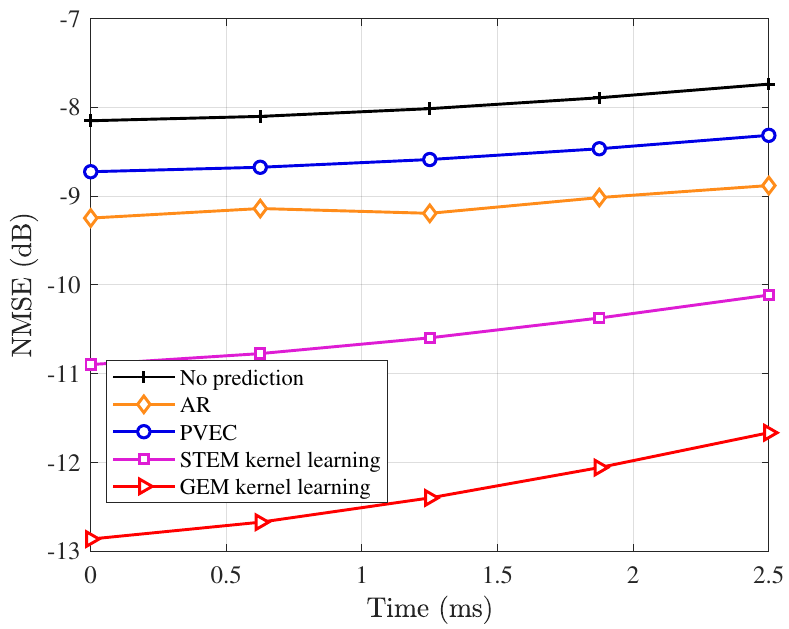}
	\caption{The NMSE performance versus time in CDL channel scenario at the maximum Doppler velocity of $36\,{\rm km/h}$.}
	\label{img: CDL_v_10_time_NMSE1}
    \vspace{-3mm}
\end{figure}

\begin{figure}[!t]
	\centering
	\setlength{\abovecaptionskip}{0.cm}
	\includegraphics[width=1\linewidth]{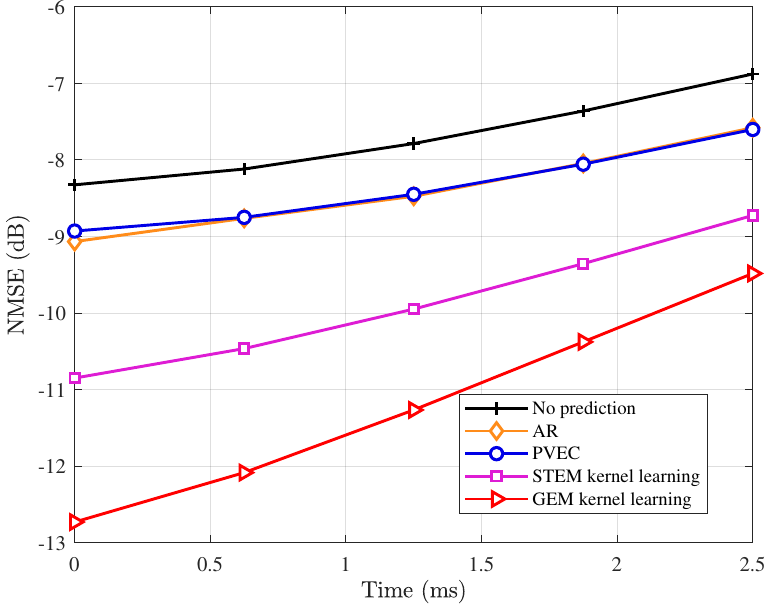}
	\caption{The NMSE performance versus time in CDL channel scenario at the maximum Doppler velocity of $72\,{\rm km/h}$.}
	\label{img: CDL_v_20_time_NMSE1}
    \vspace{-3mm}
\end{figure}

In addition, we demonstrate the temporal variation of NMSE performance corresponding to different channel prediction schemes in Fig.~\ref{img: CDL_v_10_time_NMSE1} and Fig.~\ref{img: CDL_v_20_time_NMSE1}. When ${\mathsf{SNR}}=5\,{\rm dB}$, the channels from the past two frames are used to predict the channels for the next five frames. We can observe that the proposed parallel channel prediction scheme based on GEM kernel learning also has the best NMSE performance in predicting the channels of subsequent time frames. Taking the prediction of the channel for the second future frames as an example, compared with the AR channel prediction scheme, the proposed GEM-KL method improves NMSE performance by $3.5\,{\rm dB}$ and $3.3\,{\rm dB}$ in the scenario of maximum Doppler velocity $36\,{\rm km/h}$ and $72\,{\rm km/h}$, respectively.

The above simulation results have demonstrated that on CDL channel, the channel prediction schemes based on STEM-KL can achieve better NMSE performance. It has two advantages over traditional channel prediction algorithms. On the one hand, compared to other representations of channel correlation, the EM kernel can better describe the spatiotemporal correlation of the channel. On the other hand, the STEM-KL channel prediction method can predict multiple future channels in parallel, avoiding the accumulation of errors caused by sequential prediction. The advantage of STEM with kernel learning is that it can find better hyperparameters concentration ${\bm{\delta}}$ and user motion velocity ${\bf{v}}$ for the EM kernel, which makes the STEM kernel more accurate in reflecting the spatio-temporal correlation of the channel, and therefore performs better than baseline methods. However, using gradient descent based learning methods to obtain hyperparameters relies heavily on initial values, and if the initial values are not good, the learning results may be locally optimal hyperparameters. Fortunately, the proposed GEM-KL scheme solves this problem by combining the kernels of different ${\bm{\delta}}$ and ${\bf{v}}$ grid points according to the learned optimal weights, which can avoid the problem of hyperparameter local optima and make channel prediction performance more stable. Therefore, GEM kernel learning GPR channel predictor has the best performance among all compared schemes.

\section{Conclusions}\label{sec: conclusion}
In this paper, we designed a high-accuracy channel predictor by STEM kernel learning. The STEM correlation function can capture the fundamental propagation characteristics of the wireless channel, making it suitable as a kernel function that incorporates prior information. We designed the hyperparameters of STEM kernel, including user velocity and concentration to fit time-varying channels. The hyperparameters are obtained through kernel learning. Then, the future channels are predicted through GPR, using the STEM kernel as
the prior. However, single kernel hyperparameter learning heavily relies on initial value selection. In order to further improve the stability of channel prediction, we proposed GEM-KL channel predictor. The STEM kernel is approximated by grid-based EM mixed (GEM) kernel, which is composed of STEM sub-kernels. Moreover, multi-kernel schemes are more suitable for multipath channel prediction. Finally, we conducted numerical tests on the proposed schemes using near-field multipath channel model and CDL channel model. The STEM-KL methods achieve improved performance over other baseline methods, and the GEM-KL method outperforms all compared methods. 

In future research, we will use the GEM-KL scheme to investigate other more complex channel prediction problems, such as frequency-domain wideband channel prediction.

\appendices
\footnotesize
\balance 
\bibliographystyle{IEEEtran}
\bibliography{IEEEabrv,reference}
\vspace{-1cm}
\vspace{-1cm}

\end{document}